\documentclass[prl,twocolumn,showpacs,amsmath,amssymb]{revtex4}
\usepackage{amsfonts}
\usepackage{graphicx}
\usepackage{amsmath}
\usepackage{bm}
\usepackage{amssymb}
\usepackage{soul}
\usepackage{color}
 \usepackage[version=3]{mhchem}

\begin{document}

\title{Feshbach-Stabilized Insulator of Bosons in Optical Lattices}


\author{L. de Forges de Parny$^1$, V.G. Rousseau$^2$, and T. Roscilde$^{1,3}$}  
\affiliation{$^1$ Laboratoire de Physique, CNRS UMR 5672,  \'Ecole Normale Sup\'erieure de Lyon, 
Universit\'e de Lyon, 46 All\'ee d'Italie, Lyon, F-69364, France}
\affiliation{$^2$ Department of Physics and Astronomy, Louisiana State University, Baton Rouge, Louisiana 70803, USA,}
\affiliation{$^3$ Institut Universitaire de France, 103 boulevard Saint-Michel, 75005 Paris, France}

\date{\today}

\begin{abstract}
Feshbach resonances - namely resonances between an unbound two-body state (atomic state) and a bound (molecular) state, differing in magnetic moment - are a unique tool to tune the interaction properties of ultracold atoms. Here we show that the spin-changing interactions, coherently coupling the atomic and molecular state, can act as a novel mechanism to stabilize an insulating phase  - the Feshbach insulator - for bosons in an optical lattice close to a narrow Feshbach resonance. Making use of quantum Monte Carlo simulations and mean-field theory, we show that the Feshbach insulator appears around the resonance, preventing the system from collapsing when the effective atomic scattering length becomes negative. On the atomic side of the resonance, the transition from condensate to Feshbach insulator has a characteristic first-order nature, due to the simultaneous loss of coherence in the atomic and molecular components. These features appear clearly in the ground-state phase diagram of \emph{e.g.} $^{87}$Rb around its 414 G resonance, and they are therefore directly amenable to experimental observation.
\end{abstract}


\maketitle

\emph{Introduction.} Feshbach resonances offer an invaluable tuning knob to control quantum many-body phenomena in ultracold atoms \cite{Chinetal2010}. An (unbound) state of two interacting atoms and a bound state - hereafter called molecule - are brought into resonance by the application of a magnetic field thanks to the different magnetic moments of the two states. As the two states are coupled by spin-changing interactions, their resonance allows to control the effective scattering length of unbound atoms both in magnitude and sign. This latter aspect has been widely used experimentally to explore quantum many-body phases \cite{Greineretal2003, KetterleZ2008, RanderiaT2014, Winkleretal2006, Chinetal2010} and to tune transitions between them \cite{Jordensetal08, Deissleretal2010}. On the other hand the coherent coupling between atoms and molecules has been exploited \emph{e.g.}  to observe  atom-molecule Rabi oscillations \cite{Syassenetal2007, Olsenetal2009}; at the theory level, this coupling is at the basis of the prediction of a quantum phase transition betwen mixed atom-molecule and purely molecular condensates \cite{Radzihovskyetal2004,Romansetal2004,SenguptaD2005,Radzihovskyetal2008,Ejima_2011,Bhaseenetal2012}.
 \begin{figure}
\begin{center}
\includegraphics[width=0.8 \columnwidth]{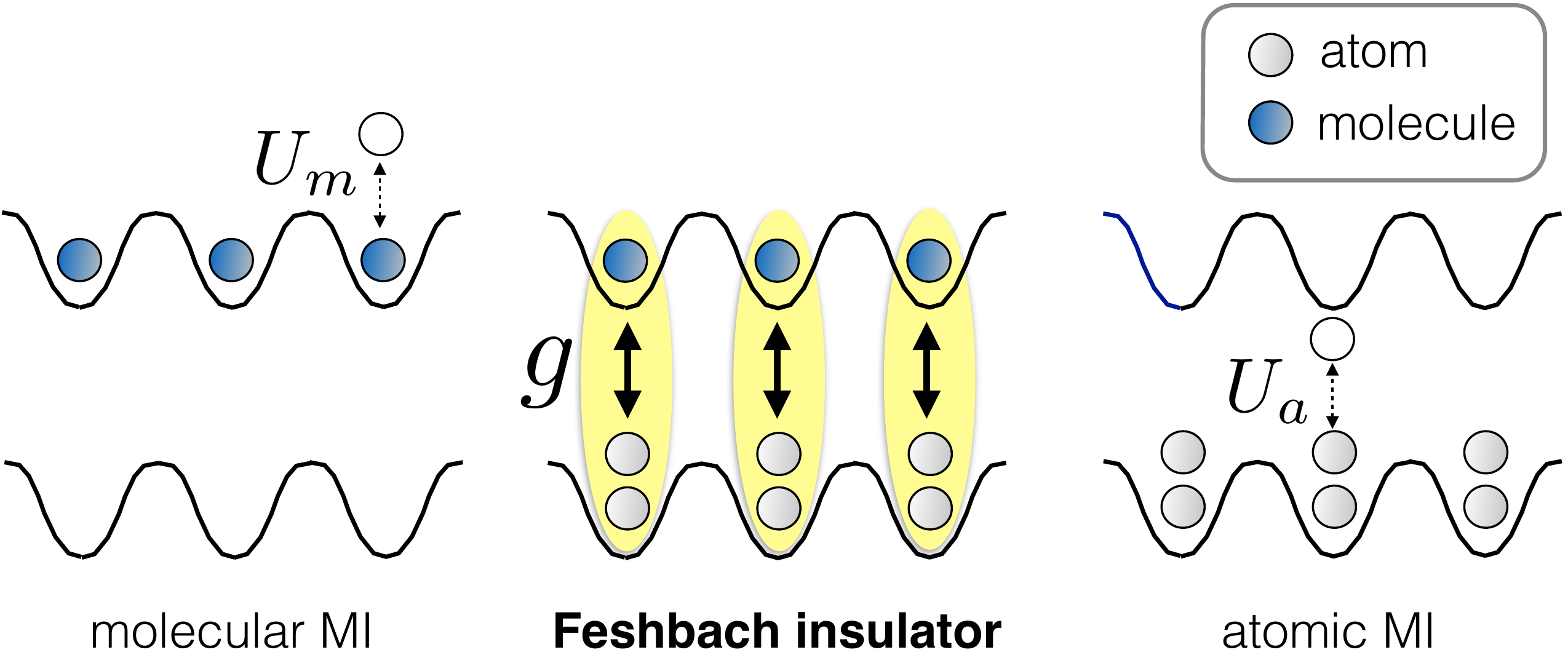}
\caption{ Various insulating phases appearing in the lattice atom-molecule coherent mixture. While atomic and molecular Mott insulators (MI) are stabilized by the repulsive interaction, the Feshbach insulator is stabilized by the atom-molecule coherence (plus a repulsive interaction to avoid collapse); yellow ellipses indicate superposition states between 2 atoms and a molecule.}
\label{fig:sketch}
\end{center}
\end{figure}

  A fundamental trait of the atom-molecule coupling is its non-linear nature: only \emph{pairs} of atoms can be converted to molecules, leading to a strong density dependence of atom-molecule conversion (AMC). Here we shall show that, in the case of bosons in optical lattices, this aspect alone can drive the system towards an insulating phase. Indeed the AMC can pin the particle density to two particles per site, opening a particle-hole (p-h) gap in the spectrum and leading to insulating behavior. We dub this state of matter a \emph{Feshbach insulator} (FI). It appears at a Feshbach resonance sufficiently narrow for the repulsive interaction to prevent the collapse of the atomic cloud. 
 AMC can be visualized as the coherent pair hopping of atoms into a fictitious, secondary lattice hosting the molecules (see Fig.~\ref{fig:sketch}). In this picture the mechanism stabilizing the FI is the appearance of strong entanglement between neighboring sites in the fictitious extra dimension, and the consequent suppression of entanglement in the real-space dimensions, which is instead at the heart of condensation. 
 
 {\color{black} A conservative definition of a FI is that of an insulating phase whose p-h gap vanishes when the AMC is suppressed; as we shall later discuss, the AMC rate is in fact tunable in the experiment, so that this characterization has a direct operational meaning. Yet the same AMC-driven mechanism, which opens a p-h gap in the FI from scratch, can be found to enhance the pre-existing gap of a molecular or atomic Mott insulator (MI); in this respect, the FI can be continuously connected to a (Feshbach-enhanced) MI. 
This picture is to be contrasted with the conventional one in which the AMC perturbatively renormalizes the atom-atom or molecule-molecule interactions via dressing the atoms (molecules) with virtual molecules (atoms): in this case p-h gaps are reduced by the AMC. We show that all three regimes of AMC (FI, Feshbach-enhanced MI and Feshbach-suppressed MI) are remarkably exhibited in the theoretical phase diagram of $^{87}$Rb close to its (narrow) 414 G resonance; moreover we discuss the condensate-FI transition, which in dimensions $D\geq 2$ is found to have a strong first-order nature.}

\textit{Atom-Molecule Hamiltonian.} $-$
We model spinless bosons in an optical lattice close to a narrow Feshbach resonance via a single-band Bose-Hubbard model \cite{footnote:H}
with atomic and molecular bosons, coherently coupled via spin-changing atom-atom interactions \cite{Koehleretal2006}. The Hamiltonian of the system reads
$\mathcal {  H}=   T+    P+  C$, where

\begin{eqnarray}
 T & = &  - \sum_{\langle i, j \rangle}  \left( t_a ~ a^\dagger_{i} a^{\phantom\dagger}_{j} + 
t_m ~m^\dagger_{i} m^{\phantom\dagger}_{j} + {\rm H.c.}  \right )\\
 P &=& \sum_i \Big[ ~\frac{U_{a}}{2}  n^a_{ i} \left ( n^a_{i}-1\right )  
+  \frac{U_{m}}{2} n^m_{i} \left ( n^m_{i} -1\right )  \\
&+ &  \nonumber U_{am} n^a_{ i} n^m_{i}
 +  (U_{a}+\delta)    n^m_{i}   - \mu   \left( n^a_{i} +  2 n^m_{i} \right)\Big]  \\
C &=&  g   \sum_{i}   \left (  m^\dagger_{i} a^{\phantom\dagger}_{i} a^{\phantom\dagger}_{i} 
+   a^{\dagger}_{i} a^{\dagger}_{i} m^{\phantom\dagger}_{i}   \right )  .
\label{termC}
\end{eqnarray}
The $ T$ operator corresponds to the kinetic energy for hopping between nearest neighboring sites $\langle i, j \rangle$ defined on a $D$-dimensional hypercubic lattice with periodic boundary conditions. 
The $a^ \dagger_{i}$ and $a^{\phantom\dagger}_{i}$ ($m^ \dagger_{i}$ and $m^{\phantom\dagger}_{i}$) operators 
are bosonic creation and annihilation operators of atoms (molecules) on site $i$. $n^a_{i}= a^{\dagger}_{i}  a^{\phantom\dagger}_{i}$ and $n^m_{i}= m^{\dagger}_{i}  m^{\phantom\dagger}_{i}$ are the corresponding number operators. 
The $ P$ operator contains the intra-species and inter-species interactions, as well as the chemical potential term; in particular it contains the detuning term $\delta$ {\color{black} (controlled experimentally by a magnetic field \cite{Chinetal2010})}, which brings the state of two atoms and a molecule in and out of resonance on each site,  $\delta<0$ ($\delta >0$) corresponding to the molecular (atomic) side of the resonance. 
Finally the $ C $ operator is the hyperfine coupling converting two atoms into a molecule and viceversa.   
          
 The above atom-molecule Hamiltonian on a lattice has been mainly studied in $D=1$ \cite{RousseauD2008, EckholtR2010, Ejima_2011, Bhaseenetal2012} for some peculiar choices of the numerous Hamiltonian parameters. Here we rather focus on the case of $D=2$ and 3 \cite{footnote:oneD}, which we investigate numerically by a combined strategy based on Gutzwiller mean-field theory (MFT) in $D=2$ and 3, supplemented in $D=2$ with numerically exact quantum Monte Carlo (QMC) simulations based on the Stochastic Green Function algorithm \cite{SGF}. MFT is found to correctly capture the succession of phases in the system,
  and it allows for the rapid reconstruction of phase diagrams. 
 The exact nature of the phase transitions encountered with MFT has also been systematically investigated {\color{black} with QMC, and will be the subject of a future publication.} 
The discussion of our results is structured as follows.  
\textcolor{black}{First, employing an idealized choice of Hamiltonian parameters, we show how the AMC term can open a p-h gap in the spectrum, giving rise to a Feshbach insulator.}
We then move on to investigating the occurrence of the FI regime for realistic parameters related to $^{87}$Rb in an optical lattice.

\begin{figure}
\begin{center}
\includegraphics[width=0.8 \columnwidth]{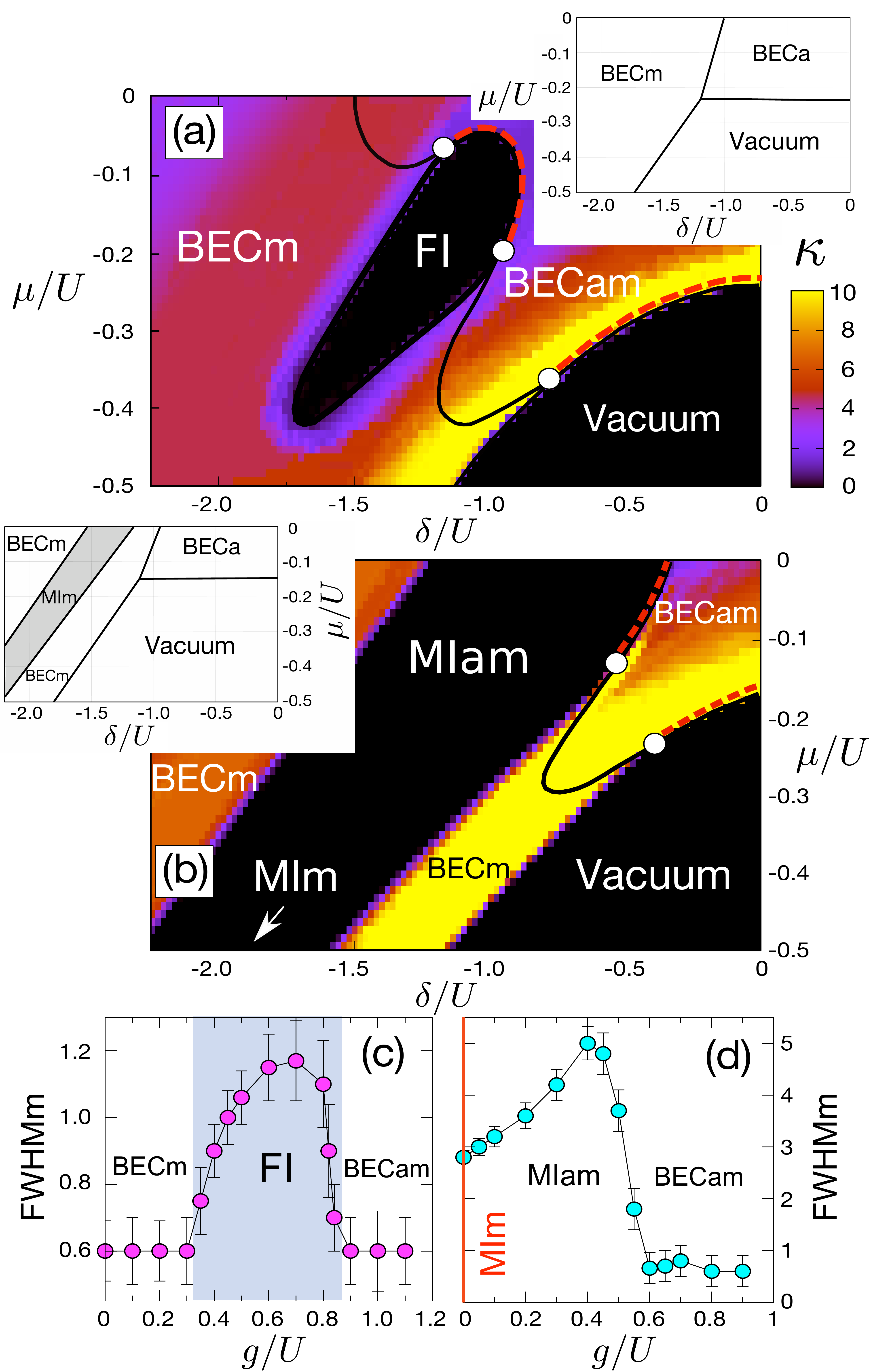}
\caption{ (a-b) Mean-field phase diagram of a symmetric atom-molecule mixture with $t/U=0.06$ and $g/U=0.8$ (a)
and  with $t/U=0.04$ and $g/U=0.6$ (b) (false colors indicate the compressibility $\kappa$). 
Inset of (a): phase diagram for $t/U=0.06$ and $g=0$. 
The following phases appear in the phase diagrams: Feshbach insulator (FI), atomic (BECa), molecular (BECm) and atomic-molecular (BECam) condensates, molecular (MIm) and 
atomic-molecular (MIam) Mott insulator. 
Second-order transitions are denoted by solid  black lines, red dashed lines indicated first-order transitions, and white dots denote tricritical points. 
(c-d) Full width at half maximum of the molecular momentum distribution (FWHMm) versus the conversion $g$ for $\delta/U=-1.2$, $t/U=0.06$  (c) and $t/U=0.04$  (d).
Resuts are from QMC simulations at fixed total density $n=2$ for a system of linear size $L=12$. 
}
\label{fig:1}
\end{center}
\end{figure}

\textcolor{black}{\textit{The Feshbach Insulator in a 2D system.} $-$}
We begin our discussion of the ground-state physics of a square lattice by imposing 
$t \equiv t_a=t_m$  and $U \equiv  U_a = U_m = U_{am} $. This establishes an (artificial) symmetry between the physics of atoms and molecules, 
\textcolor{black}{which greatly simplifies the behavior of the system and reduces the number of parameters to four only: $t/U, \delta/U, \mu/U$ and $g/U$.}


Fig.~\ref{fig:1}(a) shows the phase diagram with fixed ratio $t/U=0.06$ as a function of the detuning $\delta/U$ and of the  chemical potential $\mu/U$, as obtained via a MF calculation. In the absence of coupling $g$ between atoms and molecules (inset of Fig.~\ref{fig:1}(a)) the only phases appearing in the system are an atomic Bose-Einstein condensate (BECa), a molecular BEC (BECm) and the vacuum -- our specific choice of the $t/U$ ratio does not allow for the appearance of finite-density (Mott) insulating phases in this case (see inset of Fig.~\ref{fig:1}(a)). On the other hand, when $g>0$ ($g/U = 0.8$ in Fig.~\ref{fig:1}(a))  the phase diagram changes dramatically. The most striking feature of the phase diagram at $g>0$ is the occurrence of a broad FI insulating region at fixed total density $n = n_a + 2 n_m = 2$ \cite{footnote:resonance}. Its incompressible nature ($\kappa=dn/d\mu = 0$, see Fig.~\ref{fig:1}(a)) reveals the existence of a p-h gap which can be readily estimated from the chemical-potential width of the FI region, and which stems from the non-linear dependence of the conversion term on the occupancy \cite{footnote:gap}. 

An important ingredient for the stabilization of a homogeneous FI is the choice of a \emph{narrow} Feshbach resonance. Indeed, given that the potential energy scales as $\sim Un^2$ and the conversion energy scales as $\sim g n^{3/2}$, one finds that the two compensate each other at a density $n^{1/2} \sim g/U$. To achieve a homogeneous state, one needs therefore $g/U \lesssim n^{1/2}$ ($=\sqrt{2}$ for the example in question); if this is not the case, atoms and molecules tend to gradually cluster on single sites, with a cluster density $\sim (g/U)^2$. This clustering effect is found to initially lead to the destruction of the insulating phase and to the appearance of a simultaneous atomic/molecular BEC (BECam -- see below); but for a very large $g/U$ it ultimately leads to the collapse of the gas (see Supplementary Material (SM) \cite{Supplementary} for a detailed discussion).  

A direct experimental consequence of the p-h gap induced by the AMC in the FI is a finite coherence length. Its inverse is related to the full width at half maximum (FWHM) of the $k=0$ peak in the molecular momentum distribution $n_{m}({\bm k}) = L^{-2} \sum_{ij} e^{-i{\bm k}\cdot({\bm r}_i-{\bm r}_j)} \langle m_i^{\dagger} m_j \rangle$, as well as in the atomic one. To correctly estimate the FWHM we make use of QMC: as shown in Fig.~\ref{fig:1}(c) (calculated for a detuning $\delta/U=-1.2$), the FWHM of the molecular momentum distribution as a function of $g$ exhibits the quantum phase transition (QPT) from BECm to FI. 
Further increasing $g$ leads to a second QPT from FI to BECam \textcolor{black}{(or possibly to a very thin BECm phase  prior to entering in the BECam)}, when the conversion energy overcomes the interaction energy leading to clustering as discussed above. 
   
 When lowering the hopping to $t/U=0.04$, the phase diagram with $g=0$ (inset of Fig.~\ref{fig:1}(b)) features a molecular MI phase. As shown in Fig.~\ref{fig:1}(b), introducing a $g>0$ has the effect of enhancing the p-h gap (namely the $\mu$-width of the insulating phase), and consequently of increasing the FWHM, as shown in Fig.~\ref{fig:1}(d), giving rise to a Feshbach-stabilized MI of molecules and atoms (MIam). 
Direct inspection into the structure of the p-h excitations \cite{Supplementary} shows that the mechanism of stabilization of the MI is analogous to that opening the gap in the FI. 

 \begin{figure}
\begin{center}
\includegraphics[width=0.85 \columnwidth]{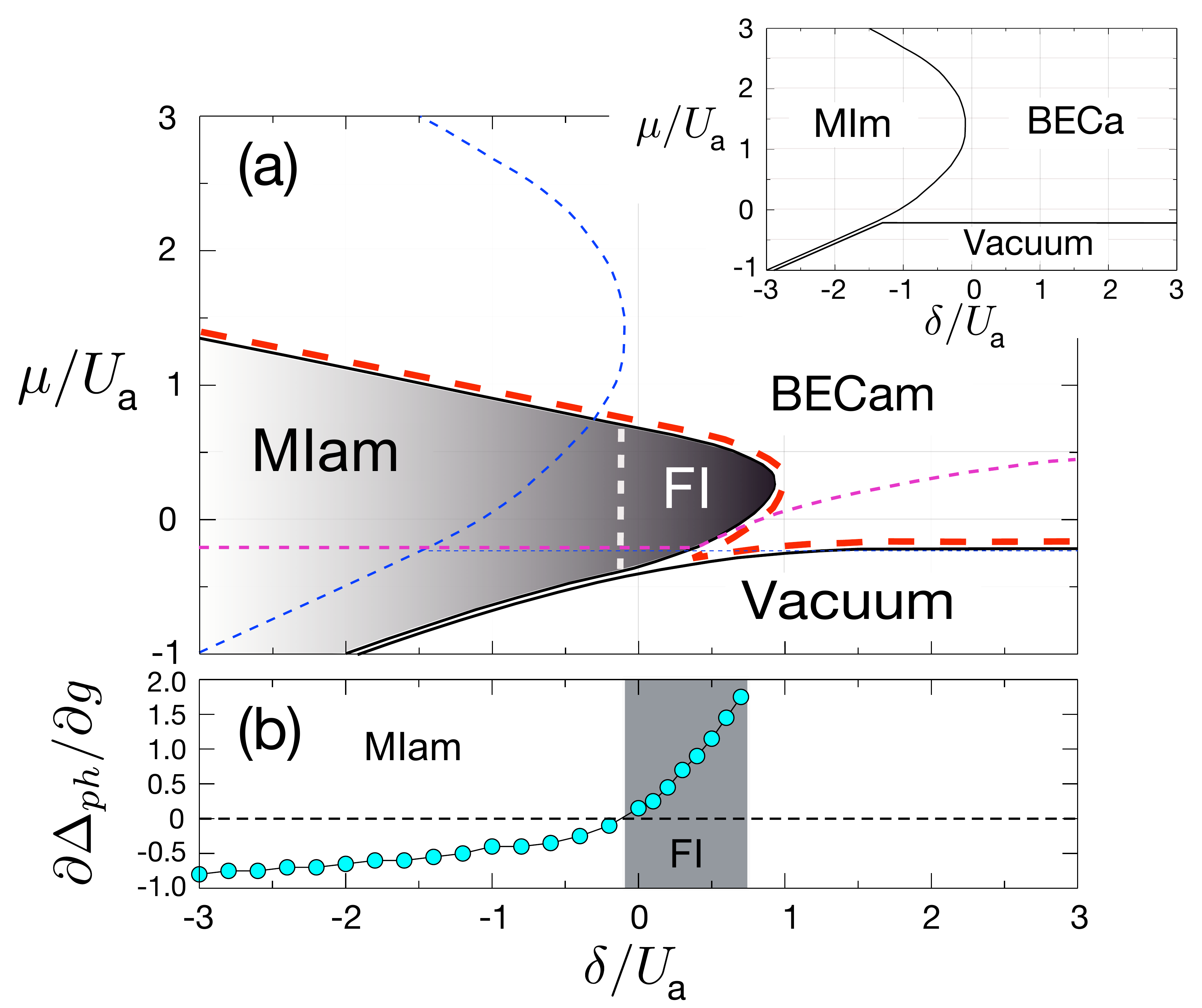}
\caption{ (a) Mean-field $T=0$ phase diagram for $^{87}$Rb close to the 414 G resonance in a cubic optical lattice of depth $V_0=12E_r$; 
The vertical dashed line marks the crossover from MIam to FI region. 
Inset: the same phase diagram obtained by artificially setting $g$ to zero. 
(b) Derivative of the p-h gap with respect to $g$: the change in sign marks the crossover from MIam to FI. 
} 
\label{fig:3}
\end{center}
\end{figure}

 \begin{figure}
\begin{center}
\includegraphics[width=1 \columnwidth]{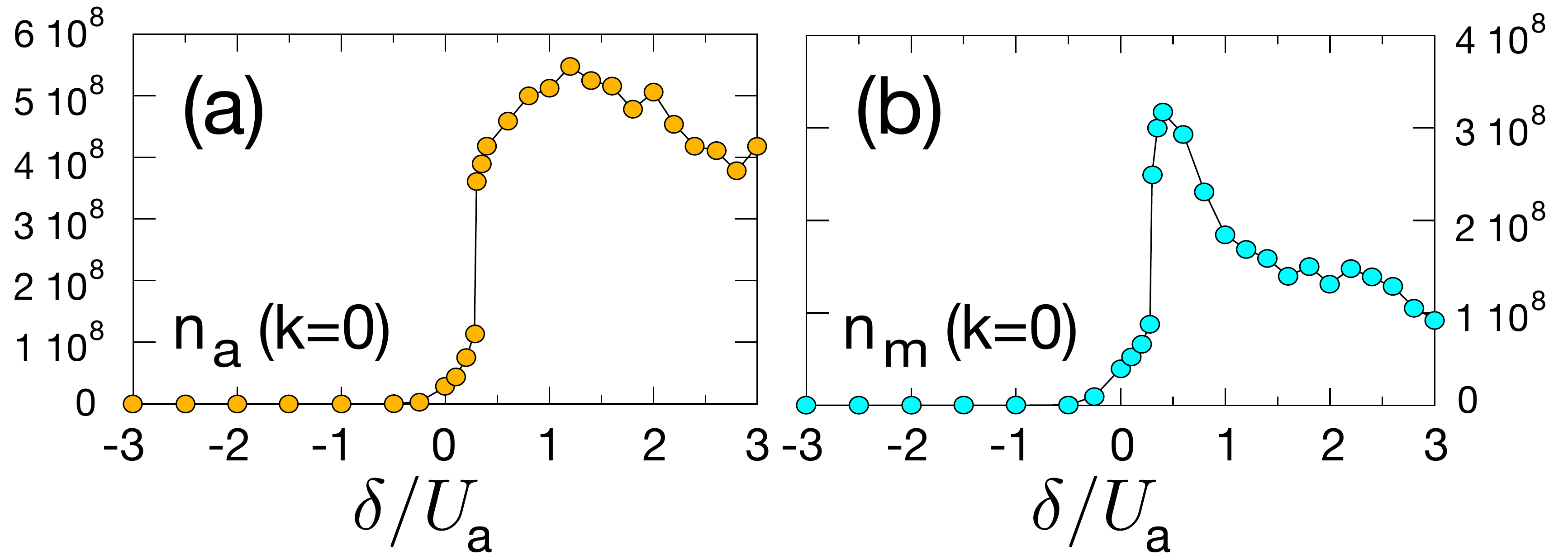}
\caption{Peaks in the atomic (a) and molecular (b) distributions for a trapped system ($\omega =  2\pi*19$ Hz) with central density $\approx 1.96$ in the BECam phase; the global chemical potential follows the dashed magenta line plot in Fig.~\ref{fig:3}(a).}
\label{fig:4}
\end{center}
\end{figure}

\textit{Phase diagram of ~$^{87}$Rb in a lattice.} $-$
We now turn our attention to a realistic implementation of the atom-molecule coherence model with a narrow Feshbach resonance, featuring most of the salient features observed in the symmetric atom-molecule mixture. To make contact with common experimental setups, we consider this time a 3D cubic optical lattice, and estimate the parameters  $t_a, t_m, U_a, U_m$ and $U_{am}$ from atomic and molecular maximally localized Wannier functions \cite{Supplementary}. The atom-molecule coherence $g$ is estimated starting from the solution of the scattering problem for two particles in a harmonic potential \cite{Buschetal1998, Syassenetal2007, Supplementary}. We have focused on the two narrow resonances at $414$ G in $^{87}$Rb, and at $853$ G for $^{23}$Na, extensively investigated experimentally 
\cite{Syassenetal2007, Inouyeetal1998, Stengeretal1999, footnote:narrowFeshbach}. The Hamiltonian parameters around these resonances have been estimated for a lattice depth $V_0 = 12 E_r$ where $E_r$ is the recoil energy \cite{footnote:recoil}.
We then scan the phase diagram as a function of chemical potential $\mu$, controlled by the trapping potential, and detuning $\delta$, controlled by the applied magnetic field, making use of MFT (expected to predict phase boundaries with an accuracy of $\sim10\%$ in 3D systems).  As the two resonances in  $^{87}$Rb and $^{23}$Na feature a similar phase diagram, in the following we shall focus on the case of $^{87}$Rb only. For $V_0 = 12 E_r$  the microscopic parameters take values $t_a/U_a= 0.04$, $t_m/U_a= 3\times 10^{-5}$, $U_m/U_a= 12.9$, $U_{am}/U_a= 3.24$ and $g/U_a= 1.23$. With this choice of the lattice depth, the atom-molecule conversion term is quite sizable ($g \approx U_a$), yet a Mott insulator of atoms cannot be stabilized at any filling (neither at the mean-field level nor at the exact level \cite{Capogrossoetal2007}): therefore insulating behavior at finite density on the atomic side of the resonance is necessarily induced by the conversion term, namely it has the nature of a FI.

 Fig.~\ref{fig:3}(a) shows the mean-field phase diagram for $^{87}$Rb at the above cited resonance. In the absence of atom-molecule conversion (inset) the system features a BECa  phase and a molecular MI (MIm) phase with $n_m=1$. 
 When $g >0$, on the other hand, the BEC phase acquires a BECam nature, and similarly the MIm acquires an atomic component (MIam).
The p-h gap $\Delta_{ph}$ of the insulating region is found to shrink substantially on the molecular ($\delta\ll \delta_c \approx -0.05 U_a$) side: this is the well-known effect of Feshbach resonances, renormalising in this case the molecule-molecule interaction -- and hence the p-h gap --- in the molecular MI via the conversion to virtual atom pairs (see \cite{footnote:nonmonotonic} for the complementary effect on the atomic side $\delta \gg \delta_c$).   
Nonetheless, for $\delta \gtrsim \delta_c$ the p-h gap is instead found to open \emph{because of} the conversion term: it vanishes in the $g\to 0$ limit \cite{Supplementary}, and it is still found to grow with the conversion rate $g$ around the value $g/U_a = 1.23$ of $^{87}$Rb. As a consequence an insulating phase is found to persist up to $\delta \approx 0.7 U_a$ on the atomic side of the resonance. Hence, even if the insulating phase on the atomic side is continuously connected with that on the molecular side, it is clear that the conversion term has an opposite role on the atomic side, opening a p-h gap instead of shrinking it. The insulating region whose p-h gap grows with $g$ has therefore the nature of a FI. Its onset is found to correspond to a change in sign of the derivative $\partial \Delta_{ph}/\partial g$, to which one can associate a change in sign of the $g$-derivative of the FWHM for the atomic momentum distribution (see  Fig.~\ref{fig:3}(b)). Both aspects can be used as a direct experimental signature -- see below. 

  As already seen in the previous example, the transition from the BECam to all the insulating phases is found to be strongly first-order. Remarkably, the unconventional nature of the BEC-insulator transition in this system can be directly observed in the experiments using state-of-the-art diagnostics. Fig.~\ref{fig:4} shows the evolution of the coherence peak across the BECam-FI transition for a trapped system with trapping frequency $\omega = 2\pi*19$ Hz, $n_{a,m}(k=0) = \frac{1}{(2\pi)^3} \sum_{ij} \phi^{*}_{i,a(m)} \phi_{j,a(m)}$~\cite{footnote:normalization}. The data for the local mean fields $\phi_{i,a(m)}$ are obtained from single-site MFT via a local density approximation. 
 We follow a trajectory in the $(\mu,\delta)$ plane - thin dashed magenta line in Fig.~\ref{fig:3}(a) - along which the density in trap center is fixed at $n \approx 1.96$ in the BECam phase -- the density jumps to $n=2$ when entering in the FI. Along this realistic trajectory in parameter space we observe a very sharp jump of both the atomic and the molecular coherent peak as the trap center crosses the BEC-insulator transition \cite{footnote:nonmonotonic}.
  A further accessible experimental evidence of the first-order nature of the BEC-insulator transition comes from the density profile. As shown in the SM \cite{Supplementary}, when the trap center is in the FI regime, the density jumps twice upon moving towards the trap wings: once when going from FI to BECam, and a second time when going from BECam to vacuum. 
    
  \emph{Conclusions.} \textcolor{black}{We have shown that coherent atom-molecule coupling at a narrow Feshbach resonance offers a novel mechanism stabilizing an insulating phase for bosons in an optical lattice -- the Feshbach insulator (FI)}. The appearance of a thermodynamically stable FI shows that an optical lattice can actively protect the bosonic cloud against the two main enemies of ultracold resonant bosons in continuum space, namely 1) collapse on the attractive side of the resonance; 2) rapid three-body recombination on the repulsive side \cite{Remetal2013,Fletcheretal2013,Makotynetal2014} -- as triple occupancy of a site is largely suppressed in the FI. How to distinguish experimentally a FI from a conventional Mott insulator? 
 As discussed above, FI behavior manifests itself for lattice depths at which atoms (and possibly even molecules) are far from a MI phase at all fillings: in this case, loss of atomic (and molecular) coherence upon tuning the system towards resonance is a clear manifestation of the appearance of a FI regime. Unlike conventional MI's, a FI consists of an almost equal mixture of atoms and molecules, $\langle n_a \rangle  \sim 2 \langle n_m \rangle$ -- an aspect directly accessible to experiments via Stern-Gerlach separation during the cloud expansion \cite{Herbigetal2003} or species-selective imaging \cite{Syassenetal2007}. 
 Atom-molecule coherence can in principle be measured on small samples via the analysis of momentum-noise correlations between atoms and molecules \cite{Supplementary}. But the most direct probe of the nature of the FI comes from the evolution of its spectral properties and coherence length upon tuning the atom-molecule coherence $g$, as shown in Fig.~\ref{fig:1}(c) and Fig.~\ref{fig:3}(b). {\color{black} The parameter $g$ can indeed be tuned below its intrinsic value via a periodic modulation of the magnetic field (already used in the experiments \cite{Thompsonetal2005} to resonantly associate molecules), which in turn drives a periodic modulation of the detuning $\delta(t) = \delta_0 + \delta_1\cos(\omega t)$. As shown in the SM \cite{Supplementary}, for $\hbar\omega, \delta_1 \ll \delta_0$ (namely far from the molecular binding energy) this leads to an effective renormalisation of $g$ to $g_{\rm eff} = g {\cal J}_0(\delta_1/\hbar\omega)$, where  ${\cal J}_0$ is the zero-th order Bessel function. This aspect paves the way for an unambiguous detection of Feshbach-stabilized insulating phases in current experimental setups.} 
  
\emph{Acknowledgements.} We thank S. D\"urr for useful discussions, and M. Eckholt-Perotti for her contributions at the early stages of this project. All calculations were performed on PSMN cluster (ENS Lyon). This work is supported by the ANR-JCJC programme (ArtiQ project).

\pagebreak
\newpage

\begin{center}
\section{Supplementary Material: Feshbach-Stabilized Insulator of Bosons in Optical Lattices}
\end{center}

\maketitle

\section{1. Methods used}
In $D=2$ and 3 we supplement the QMC calculations with Gutzwiller mean-field theory (MFT), consisting in solving the problem of a single site coupled to the self-consistent atomic and molecular mean fields, $\phi_a = \langle a \rangle$ and 
 $\phi_m = \langle m \rangle$, respectively.  Moreover in $D=2$, we make use of numerically exact quantum Monte Carlo (QMC) simulations based on the Stochastic Green Function (SGF) algorithm with directed updates \cite{SM_SGF}, both in a canonical and grand-canonical setting. We treat $L\times L$ lattices with sizes up to $L=14$. An inverse temperature of $\beta t=2L$ {\color{black} (where $t=\min(t_a,t_m)$)} allows to eliminate thermal effects from the QMC results. 
 
 When comparison between MFT and QMC is made, MFT is found to correctly capture the succession of phases in the system upon varying the Hamiltonian parameters, although the agreement between the two approaches is typically semi-quantitative, as it will be further discussed below.

\section{2. Additional information on the mean-field phase diagram}

In this section, we provide additional elements concerning the mean-field phase diagram plot in Fig.~\ref{fig:1}(a)
of the main text. 
The phase diagram is obtained by self-consistent minimisation of the following single-site Hamiltonian
\begin{eqnarray}
 {\cal H}_{\rm MF} & = &  -  \left( z t_a ~ \phi_a a^\dagger +
z t_m \phi_m ~m^\dagger + {\rm H.c.}  \right)\nonumber \\
&+& z t_a ~ |\phi_a|^2+z t_m ~ |\phi_m|^2 \nonumber \\
 &+&   \frac{U_{a}}{2}~  n^a \left (n^a-1\right )  
+  \frac{U_{m}}{2} ~n^m \left ( n^m -1\right )  \nonumber \\
&+ &  \nonumber U_{am} ~n^a n^m
 +  (U_{a}+\delta)~    n^m   - \mu ~ \left( n^a +  2 n^m \right)  \nonumber \\
 &+& g   \left (  m^\dagger a^{\phantom\dagger} a^{\phantom\dagger} 
+   a^{\dagger} a^{\dagger} m^{\phantom\dagger}   \right )  
\label{e.MF}
\end{eqnarray}
where $\phi_a = \langle a \rangle$ and $\phi_m = \langle m \rangle$ are evaluated on the ground state of ${\cal H}_{\rm MF}$; $z$ is the coordination number. 

\begin{figure}[h!]
\begin{center}
\includegraphics[width=1 \columnwidth]{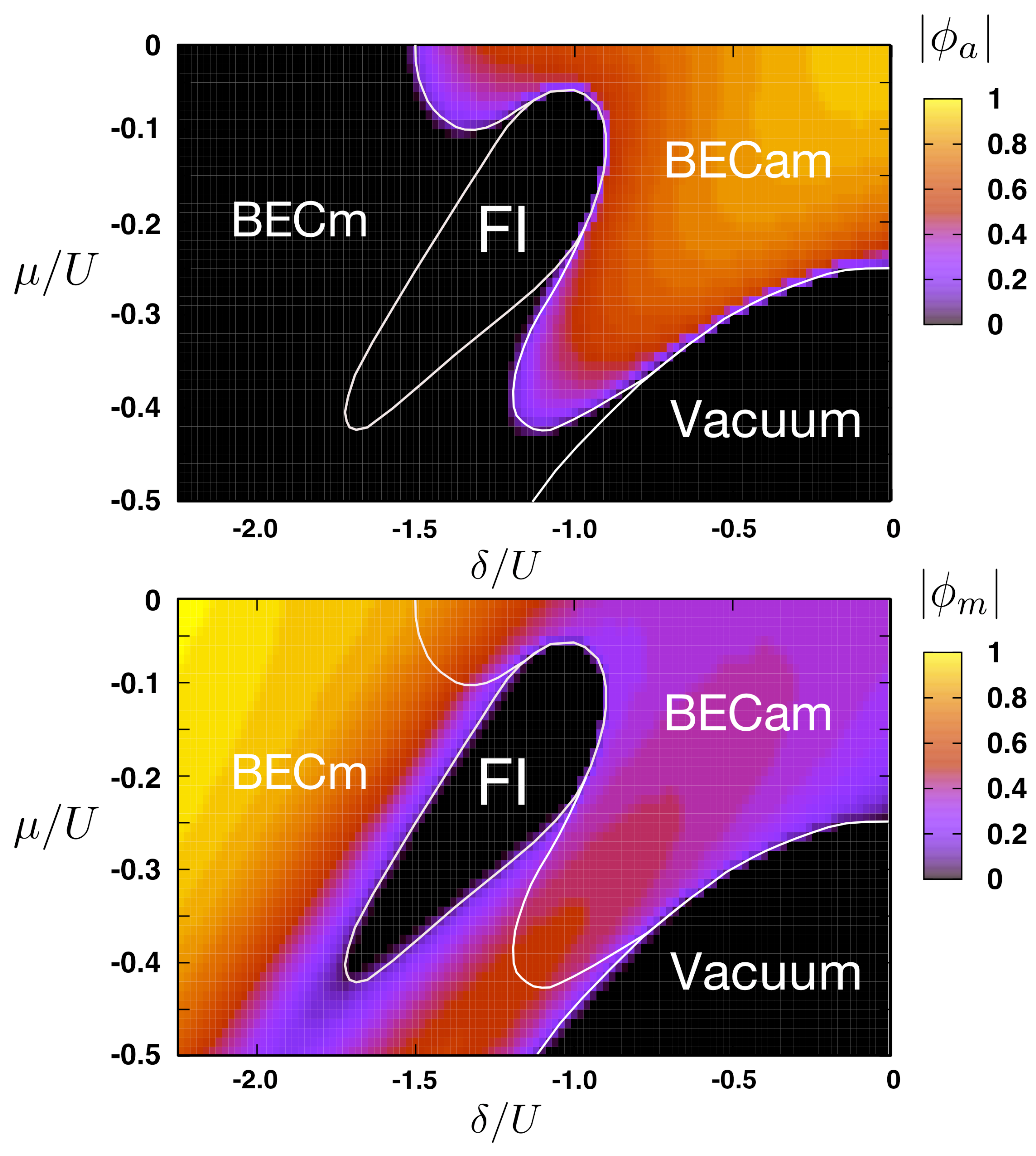}
\caption{ Modulus of the atomic (up) and molecular (down) coherences from the mean-field phase diagram plot in 
Fig.~\ref{fig:1}(a) of the main text. 
The atomic coherence vanishes continuously  at the BECam to BECm transition (upper panel) whereas the molecular coherence remains finite (lower panel).
Both atomic and molecular coherences vanishe in the Feshbach insulator (FI) phase.}
\label{phiaphim}
\end{center}
\end{figure}

\begin{figure}[h!]
\begin{center}
\includegraphics[width=0.95 \columnwidth]{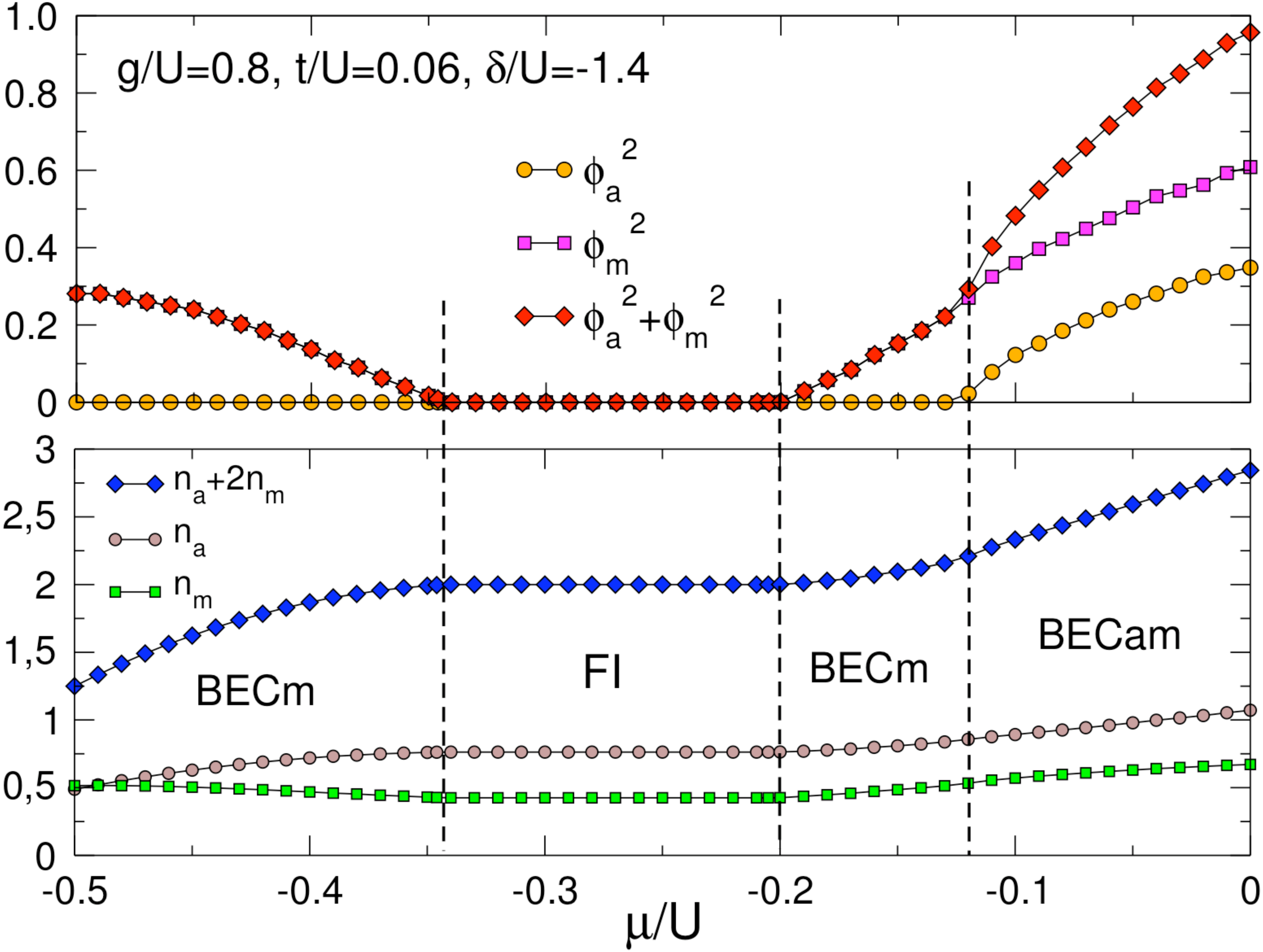}
\includegraphics[width= 0.95 \columnwidth]{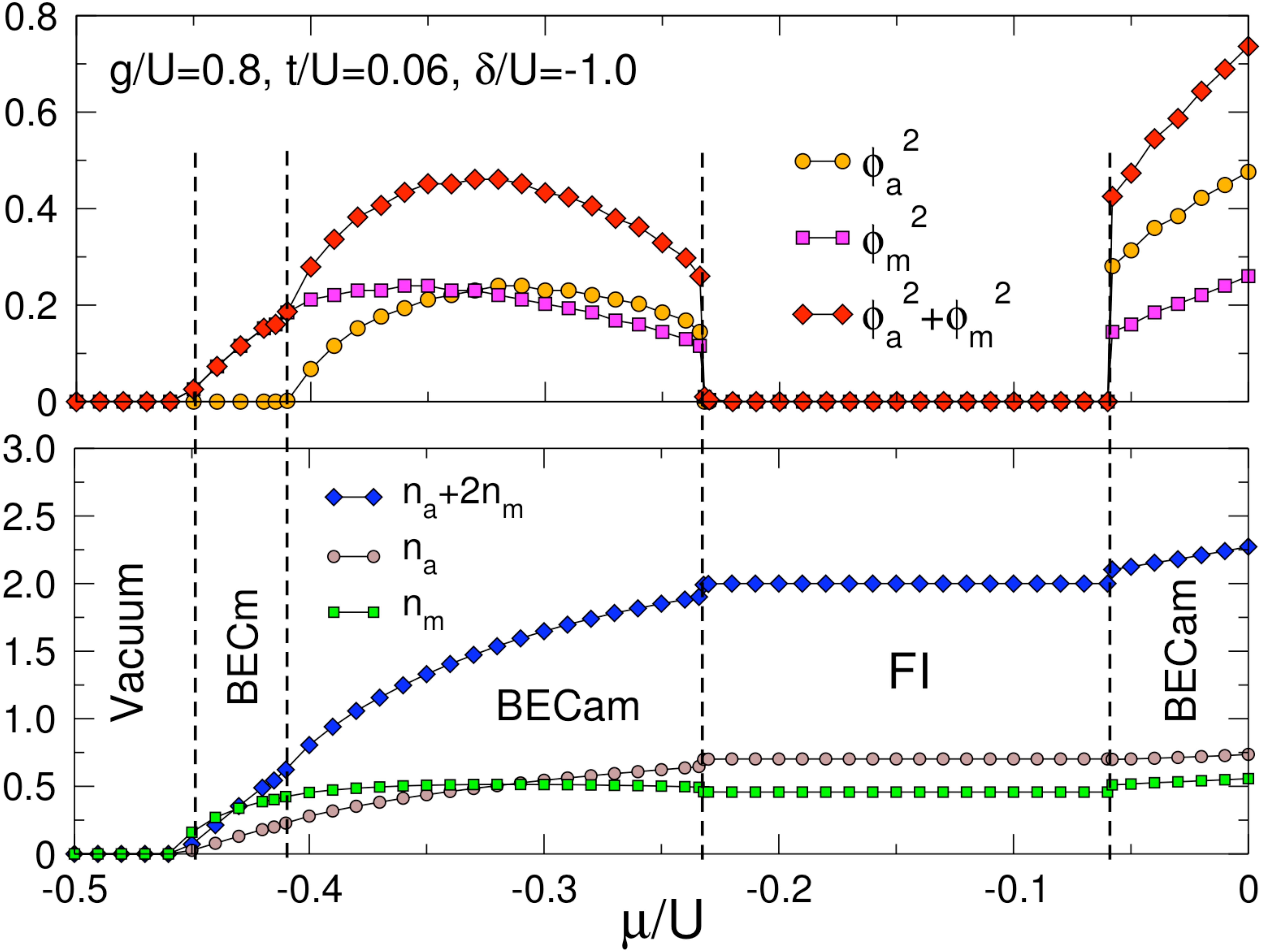}
\caption{ (a) Vertical cuts of the mean field phase diagram (Fig.~\ref{fig:1}(a)  of the main text) for $\delta/U=-1.4$ (top) and $\delta/U=-1.0$ (bottom).
All the transition are continuous for $\delta/U=-1.4$, whereas the atomic-molecular BEC (BECam) to the Feshbach insulator (FI) is first order for  $\delta/U=-1.0$.}
\label{Vertical_slices_mean_field_phase_diagram}
\end{center}
\end{figure}

Fig.~\ref{phiaphim} shows the modulus of the atomic and molecular coherences ($\phi_a$ and $\phi_m$), exhibiting the presence of two distinct Bose-Einstein condensate (BEC) phase: atom-molecule BEC (BECam) and molecular BEC (BECm), separated by a continuous transition. Moreover two insulating phases are present: the vacuum phase and the Feshbach insulator (FI).  
Vertical cuts through this diagram, obtained by increasing $\mu/U$ for two different values of $\delta/U$,  are shown in Fig.~\ref{Vertical_slices_mean_field_phase_diagram}. The cut at $\delta/U=-1.4$ shows the continuous transitions between BECm and FI both for the atomic/molecular coherences as well as for the densities. On the other hand, the cut at $\delta/U=-1$ shows that the BECam-FI transition has a first-order nature, with clear jumps in both coherences and densities.  

\begin{figure}[ht!]
\begin{center}
\includegraphics[width=7 cm]{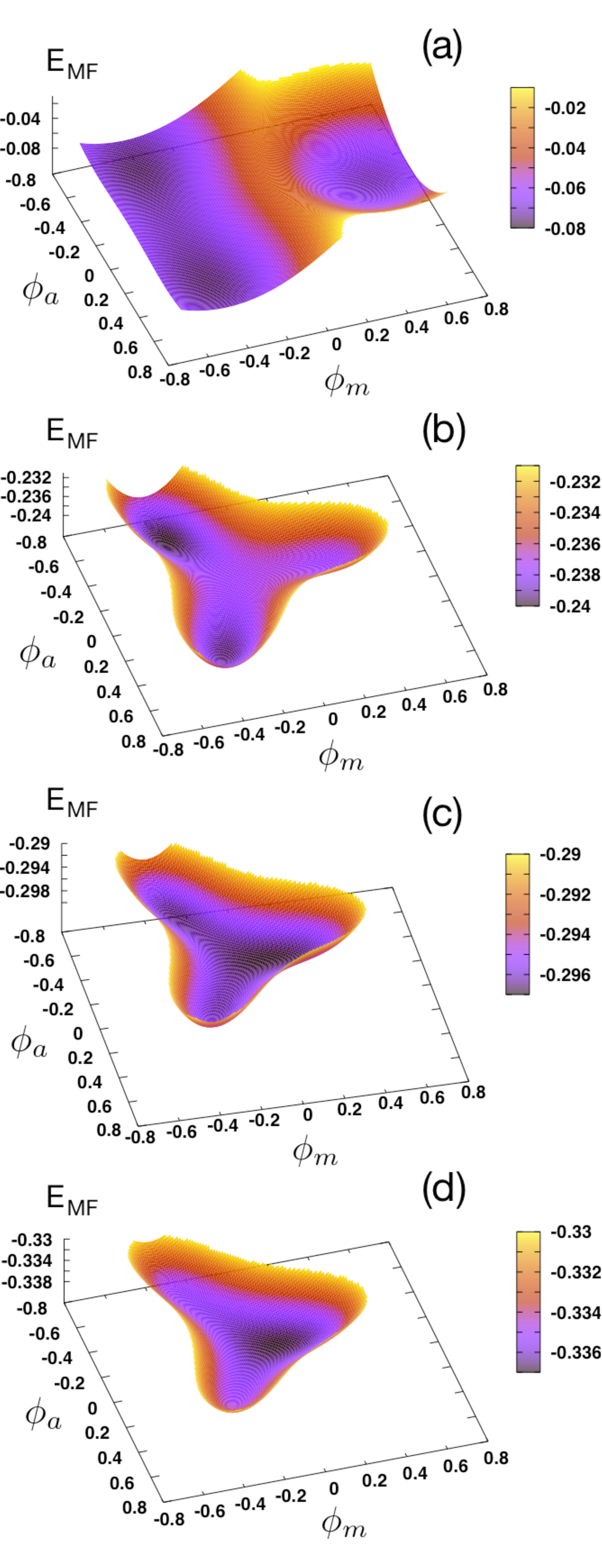}
\caption{ Mean field ground-state energy $E_{\rm MF}$
with $g/U=0.8$, $t/U=0.06$ and $\delta/U=-1.0$.
(a) $\mu/U=-0.35$ (BECam);
(b) $\mu/U=-0.25$ (BECam);
(c) $\mu/U=-0.22$ (FI);   
(d) $\mu/U=-0.20$ (FI).}
\label{energy_landscape_first_order}
\end{center}
\end{figure}

The first-order transition implies a coexistence between different phases. This can be clearly singled out at the mean-field level by setting $\phi_a$ and $\phi_m$ in the mean-field Hamiltonian Eq.~\eqref{e.MF} as parameters, and reconstructing the ground-state mean-field energy function $E_{\rm MF}(\phi_a,\phi_m)$ (both $\phi_a$ and $\phi_m$ are assumed to be real without loss of generality). Fig.~\ref{energy_landscape_first_order} shows the mean-field ground-state energy $E_{\rm MF}$ as $\mu/U$ is increased across the BECam-FI transition at $\mu/U \simeq -0.23$ and $\delta/U=-1.0$. For $\mu/U=-0.35$ 
(Fig.~\ref{energy_landscape_first_order}(a)) the energy landscape clearly shows three minima: two stables ones with $\phi_a\neq 0$ and $\phi_m\neq 0$ (corresponding to a BECam ground state), and a metastable one with $\phi_a= 0$ and $\phi_m\neq 0$  (corresponding to a metastable BECm). The three minima exhibit the strong asymmetry present between atoms and molecules due to the atom-molecule conversion term: when the phase of the molecular field is fixed (to zero in this case), the mean-field energy remains symmetric under the transformation $\phi_a \to -\phi_a$, manifesting a residual $\mathbb{Z}_2$ symmetry in the choice of the phase of the atomic field. As $\mu/U$ increases, the metastable BECm minimum is found to jump to the origin $\phi_a = \phi_m = 0$, acquiring the nature of a FI (\textit{e.g.} Fig.~\ref{energy_landscape_first_order}(c) and (d)), and to become degenerate with the BECam minima for $\mu/U=-0.23$. The presence of multiple minima at the transition point is characteristic of a first-order transition. 

\begin{figure}[h!]
\begin{center}
\includegraphics[width=0.9 \columnwidth]{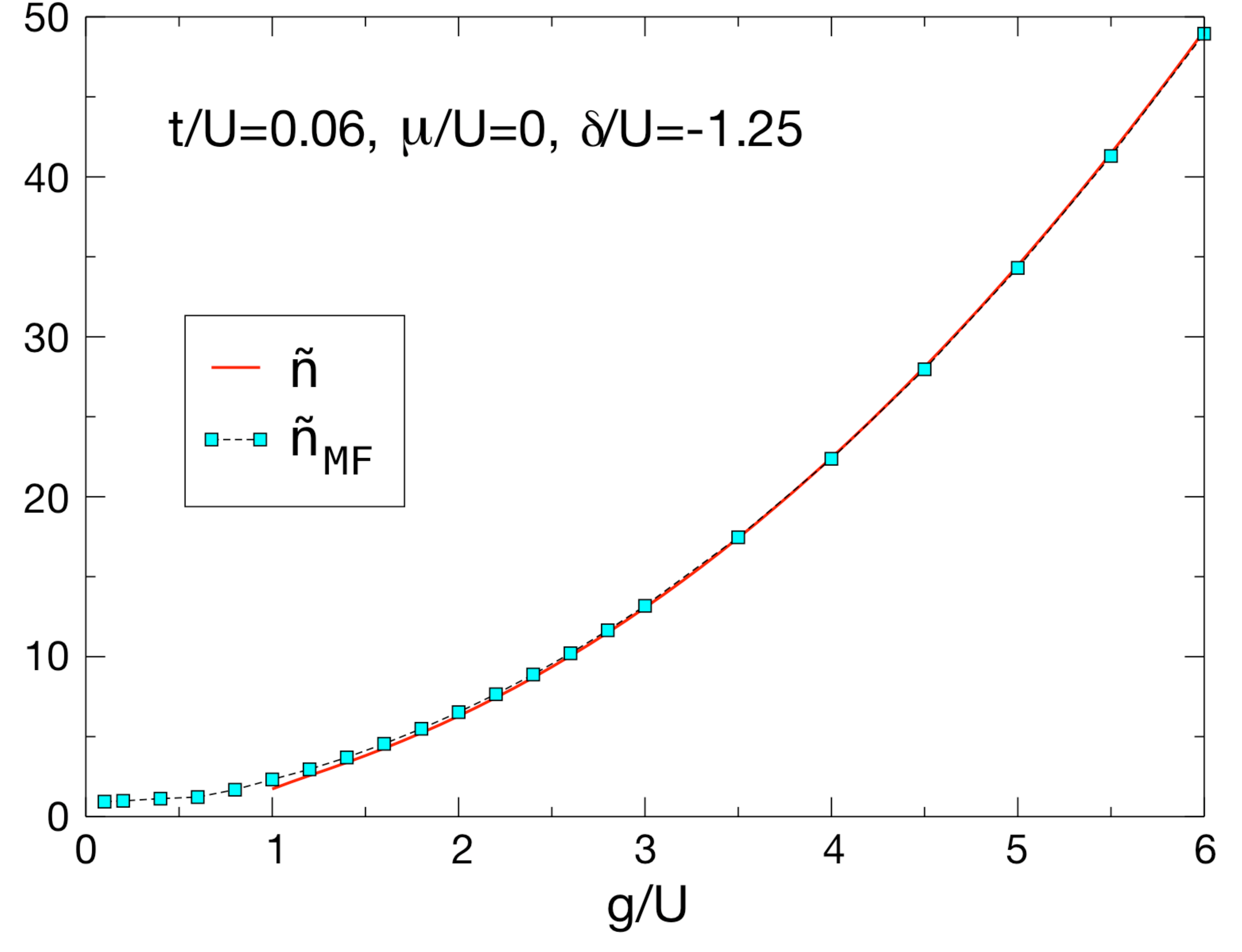}
\caption{ Total number density as a function of $g/U$ for the 2D atom-molecule symmetric model ($t_a = t_m = t$, $U_a = U_m = U_{am} = U$) with $t/U = 0.06$, $\delta/U = -1.25$, and $\mu=0$. 
The squares indicate the full mean-field solution ${\tilde n}_{MF}$ (from the minimization of the Hamiltonian Eq.~\eqref{e.MF}), while the solid line is the prediction from the approximate energy function, Eq.~\eqref{e.approx}.}
\label{fig:n_increasing_g}
\end{center}
\end{figure}

\begin{figure}[h!]
\begin{center}
\includegraphics[width=1 \columnwidth]{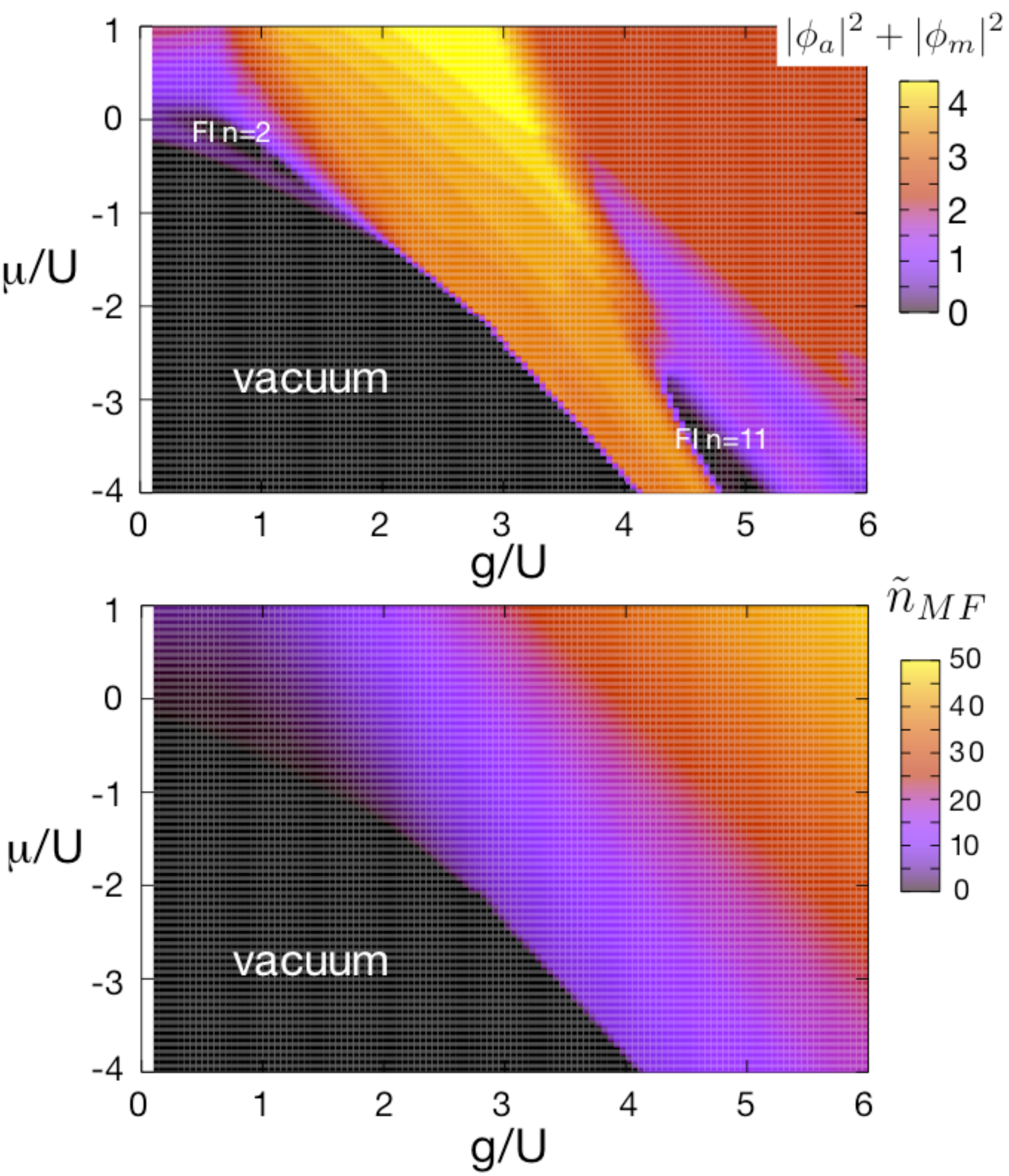}
\caption{ Mean-field phase diagram at $T=0$ for the 2D atom-molecule symmetric model ($t_a = t_m = t$, $U_a = U_m = U_{am} = U$) with $t/U = 0.06$, $\delta/U = -1.25$ upon varying the chemical potential $\mu$ and the atom-molecule coupling $g$. False colours indicate the total coherent fraction (upper panel) and the total mean-field number density (lower panel).}
\label{fig:phd_increasing_g}
\end{center}
\end{figure}

\section{3. From Feshbach insulator to collapse for broad resonances}
\textcolor{black}{
As mentioned in the main text, increasing the strength $g$ of the atom-molecule coupling -- controlled by the width of the Feshbach resonance -- one expects to gradually destabilize the system towards collapse. In particular, a simple scaling argument predicts that the total density increases as $n \sim (g/U)^2$ when $g$ increases. In the case of the symmetric atom-molecule mixture ($t_a = t_m = t$, $U_a = U_m = U_{am} = U$) one can conduct a more refined analysis based on an approximate mean-field energy function, uniquely dependent upon the atomic and molecular densities $n_a$ and $n_m$, and taking the form
\begin{eqnarray}
{\cal E}(n_a,n_m) &=&  \frac{U}{2} \left[  n_a(n_a-1) + n_m(n_m-1) + 2 n_a n_m \right] \nonumber \\
&-& 2 g n_a \sqrt{n_m} - \mu (n_a + 2n_m) + (\delta+U) n_m. \ \ \ \  \ 
\end{eqnarray}
Here we have neglected the kinetic energy altogether - this is justified in the limit in which $g$ and $U$ are the largely dominant energy scales in the system. Moreover, in the conversion term we have assumed a phase difference of $\pi$ between the atom pairs and the molecules, minimizing the conversion energy. In the case $\mu = 0$, minimization of the above density function with respect to $n_a$ and $n_m$ leads to a rather simple result for the equilibrium value of the total \emph{number} density ${\tilde n} = n_a + n_m$
\begin{eqnarray}
{\tilde n} & = & \frac{2}{3} \left( \frac{1}{4} + \left ( \frac{g}{U} \right )^2 - \frac{\delta}{2U} \right) \nonumber \\
& + & \sqrt{ \frac{4}{9} \left( \frac{1}{4} +
  \left( \frac{g}{U} \right)^2 -  \frac{\delta}{2U} \right)^2 + \frac{1}{3} \left( 1+ \frac{4\delta}{U} \right) }
\label{e.approx}  
\end{eqnarray}
exhibiting again the anticipated $(g/U)^2$ scaling. As shown in Fig.~\ref{fig:n_increasing_g} the above result reproduces very well the full solution of the self-consistent minimization of the mean-field energy in the case $t/U = 0.06$ and $\delta/U = -1.25$. }

\textcolor{black}{
Fig.~\ref{fig:phd_increasing_g} shows the mean-field phase diagram of the symmetric atom-molecule mixture with $t/U = 0.06$ at fixed $\delta/U = -1.25$ and variable chemical potential and atom-molecule coupling, reconstructed via the total coherent fraction $|\phi_a|^2 + |\phi_m|^2$ and the total mean-field number density ${\tilde n}_{MF} = n_{a} + n_{m}$. The FI for $n=2$ is found to persist in the phase diagram only for moderate values of $g/U (\sim 1.5)$, beyond which it leaves space to an atomic-molecular condensate phase with increasing density. Interestingly, another incompressible insulating phase is found at (mass) density $n=11$ for the particular value of the detuning chosen here - suggesting that in fact a whole family of FIs of increasing density might be stabilized in the system - yet a full characterization of the phase diagram upon varying the detuning $\delta$ goes beyond our current scopes. }

\section{4. Hamiltonian parameters for a atom-molecule resonant mixture in a lattice}

The Hamiltonian parameters $t_a, t_m, U_a, U_m, U_{am}$ for $^{87}$Rb atoms and  $^{87}$Rb$_2$ Feshbach molecules in a cubic optical lattice, with wavevector $k$ and depth $V_0$, are estimated making use of the numerically calculated band structure and maximally localized Wannier states \cite{SM_WannierFunction}. The atom and molecule tunnelling matrix element $t_a$ and $t_m$ 
are directly extracted from the width of the lowest energy band. The molecules see an optical lattice of effective depth $4V_0$, due to the fact that the dipole matrix element, whose square enters $V_0$, takes contribution from both atoms of the molecule. This leads to  $t_a \gg t_m$ (by orders of magnitude), and not to $t_a = 2 t_m$, as often assumed in the literature.

The on-site atom-atom, molecule-molecule and atom-molecule interactions matrix element, respectively
 $U_a$, $U_m$ and $U_{am}$, are obtained as 
\begin{eqnarray}
U_a&=&\frac{4\pi \hbar^2 a_{bg}}{m} \int d^3 \textbf{r}  |w_{a} (\textbf{r} ) |^4 \\
\label{terme_Uaa}
U_m&=&\frac{16\pi \hbar^2 a_{bg}}{m} \int d^3 \textbf{r}  |w_{m} (\textbf{r} ) |^4 , \\
\label{terme_Umm}
U_{am}&=&\frac{8\pi \hbar^2 a_{bg}}{m} \int d^3 \textbf{r}  |w_{a} (\textbf{r} ) |^2  |w_{m} (\textbf{r} ) |^2 ,
\label{terme_Uam}
\end{eqnarray}
with $a_{bg}$ being the background scattering length of the atoms, $m$ the atomic mass  and 
$w_{a (m)}(\textbf {r})$ the atomic (molecular) Wannier function.

The conversion rate between atoms and molecules $g$ is obtained via the solution of the scattering problem for two atoms in a parabolic potential \cite{SM_Buschetal1998}. Following \cite{SM_Syassenetal2007}, the parameter $g$ is given by
\begin{eqnarray}
 \frac{g}{U_{a,h}} =~~~~~~~~~~~~~~~~~~~~~~~~~~~~~~~~~~~~~~~~~~~~~~~~~~~~~~~~~~~~~ && \nonumber \\
 \left [ \frac{\sqrt{\pi} m \Delta \mu \Delta B }{\sqrt{2} \hbar^2 k^3 a_{bg} }
 \left( \left ( \frac{V_0}{E_r} \right )^{-\frac{3}{4}}+0.49 k a_{bg} \left (\frac {V_0}{E_r} \right)^{-\frac{1}{2}} \right ) \right ]^{\frac{1}{2}} 
\label{terme_g}
\end{eqnarray}
with $\Delta \mu$ the difference of magnetic moments, $\Delta B$ the width of the Feshbach resonance, 
$V_0$ the lattice depth seen by an atom and $E_r=\hbar^2 k^2/2m$ the atomic recoil energy.
Here {$g$} is expressed in units of $U_{a,h}$, namely the atom-atom repulsion within the harmonic approximation for the Wannier function,  $U_{a,h}=\sqrt{8/\pi}~k a_{bg} (V_0/E_r)^ \frac{3}{4} E_r $. For the lattice depth we considered ($V_0=12 E_r$) $U_{a,h}$ overestimates $U_a$ by some 20\%. Yet this choice of normalization for $g$ is justified by the fact that $g$ is also calculated by assuming a single harmonic potential, and hence can be expected to be similarly overestimated with respect to the actual value. 

\textcolor{black}{
 Making contact with the previous section, it is interesting to summarize here our estimates for the value of the $g/U_{a,h}$ ratio associated with some relevant narrow Feshbach resonances already investigated experimentally \cite{SM_Chinetal2010}. Considering optical lattices of depth $V_0$ varying between 4 and $30~ E_r$, we find
 \begin{enumerate} 
\item $^{87}$Rb, $B_0=414$ G: $g/U_{a,h} \sim 2-1$; 
\item $^{23}$Na, $B_0=853$ G: $g/U_{a,h} \sim 3-1$;
\item $^{87}$Rb, $B_0=9$ G: $g/U_{a,h} \sim 8.5-4$;
\item ~$^{6}$Li, \ \ $B_0=543$ G: $g/U_{a,h} \sim 7-3.5$.
\item $^{87}$Rb, $B_0=1007$ G: $g/U_{a,h} \sim 40-20$; 
\item $^{23}$Na, $B_0=907$ G: $g/U_{a,h} \sim 60-30$; 
\end{enumerate}
(Here $B_0$ indicates the position of the resonance in magnetic field). 
As already mentioned in the main text, apart from the first two resonances, we cannot find a stable, incompressible FI at density $n=2$. This is consistent with the results of the previous section, which, albeit focusing on an artificially symmetric atom-molecule mixture, were also suggesting that a $n=2$ FI is only expected for $g/U \sim 1$. }

\begin{figure}[ht!]
\begin{center}
\includegraphics[width=0.9 \columnwidth]{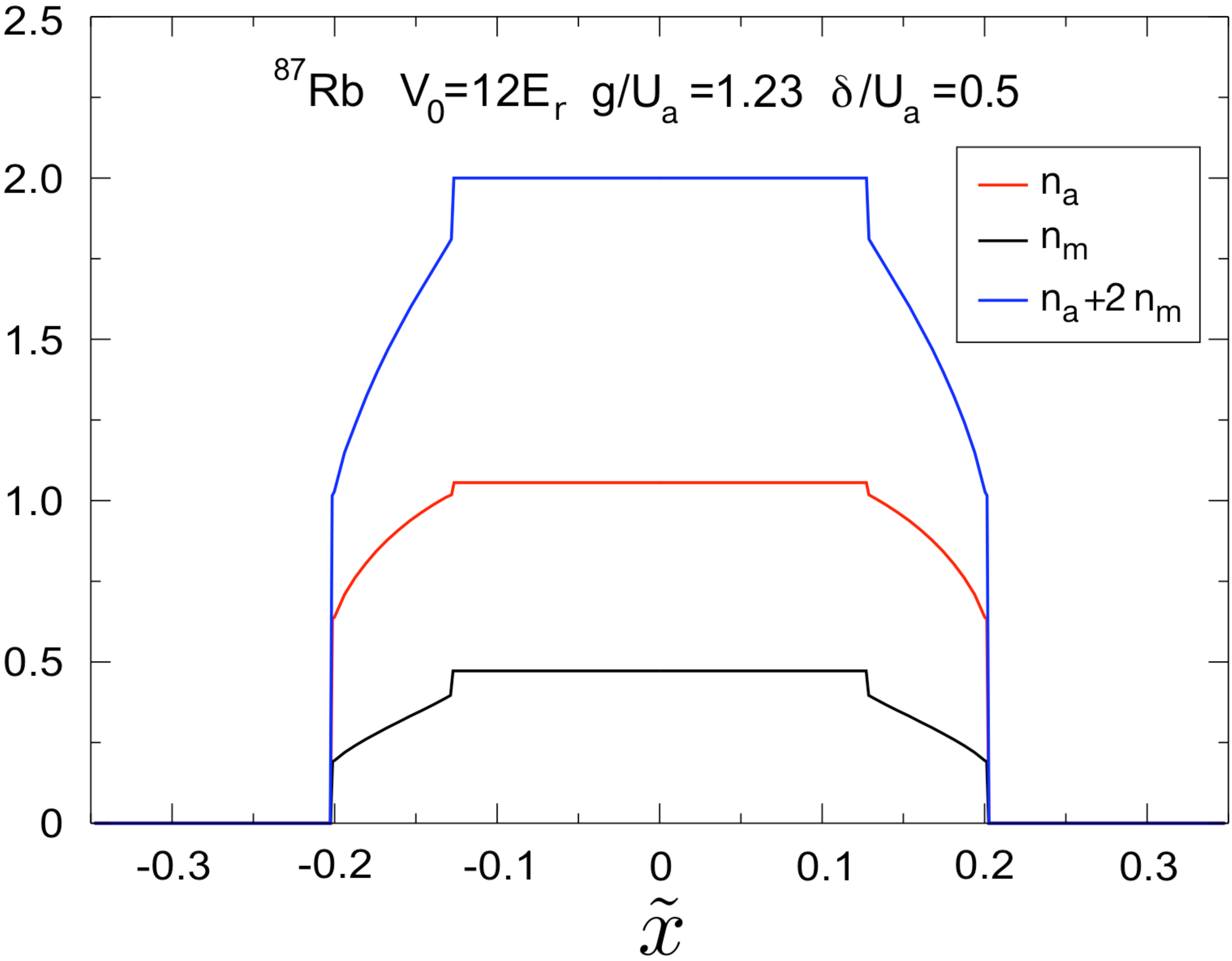}
\caption{ Atomic, molecular and total density profile for $^{87}$Rb in a trap plus cubic optical lattice, with $\mu/U_a=0.1$ and $\delta/U_a=0.5$; other parameters as in 
Fig.~\ref{fig:3}(a) of the main text.
The abscissa represents the scaled radial variable $\tilde x = V_t/U_a (r/d)^2$, where $d$ is the lattice spacing, $r$ is the radial distance from the trap center, and $V_t = (1/2)m\omega^2 d^2$ is the trapping potential.}
\label{fig:SM_Figure6}
\end{center}
\end{figure}

\section{5. Density profiles for near-resonant $^{87}$Rb in a trap}

 Fig.~\ref{fig:SM_Figure6} shows the density profile of $^{87}$Rb in a cubic optical lattice of depth $V_0/E_r = 12$ plus a harmonic trap, and in a magnetic field close to its 414-G Feshbach resonance. The chemical potential and detuning take values $\mu/U_a=0.1$ and $\delta/U_a=0.5$, which set the trap center into a gapped FI phase with total density $n=2$. The resulting density profile is obtained using standard local-density approximation. Moving towards the trap wings, we observe a density jump in both the atomic and molecular components when the system goes locally from FI to BECam, and a further jump when going from BECam to vacuum. Such jumps are observed as well for square lattices, and they are therefore directly amenable to experimental observation using \emph{e.g} quantum gas microscopes \cite{SM_Bakretal2009}.

\section{6. Controlling the particle-hole gap with the atom-molecule conversion}

In this section we discuss how the role of the atom-molecule conversion rate $g$ in opening, stabilizing or suppressing the particle-hole (p-h) gap in the system can be fundamentally understood from the structure of the particle and hole excitations around a given filling, whose wave function can be extracted from MFT. This analysis allows to contrast the various cases of FI, Feshbach-enhanced MI or Feshbach-suppressed MI encountered in the main text.

We first focus on the symmetric atom-molecule mixture (Fig.~\ref{fig:1} of the main text). In this case, we show that the atom-molecule conversion may open a p-h gap, driving a quantum phase transition from a BEC to a FI; or enhance the p-h gap of a pre-existing MI for sufficiently small $g$. In both cases, for larger values of $g$ the atom-molecule conversion has instead the opposite role of suppressing completely the p-h gap. 
Then, we shall focus our attention on the case of $^{87}$Rb: for this system the distinction between Feshbach and (molecular) Mott insulator is particularly important as the two phases are continuously connected upon varying the detuning (Fig.~\ref{fig:3}(a) of the main text). 
The analysis of the nature of the particle-hole gap provides further details concerning the crossover between the two regimes -- beyond the sign change in the derivative of the p-h gap with respect to $g$, discussed in the main text (Fig.~\ref{fig:3}(b)).

\subsection{Symmetric atom-molecule mixture}
 
 \subsubsection{Dependence of the particle-hole gap on conversion rate}

Focusing on the symmetric case ($t \equiv t_a=t_m$  and $U \equiv  U_a = U_m = U_{am} $), we discuss the p-h gap evolution versus $g$ for a fixed detunning $\delta/U=-1.2$ and for the two values of the hopping, $t/U=0.04$ and $t/U=0.06$,  discussed in the main text. As shown in Fig.~\ref{fig:1} of the main text, at this value of the detuning for $g=0$ a Mott phase (of molecules, MIm) exists only for $t/U=0.04$, whereas the system is in a molecular BEC phase (BECm) for $t/U=0.06$. 

\begin{figure}[ht!]
\begin{center}
\includegraphics[width=\columnwidth]{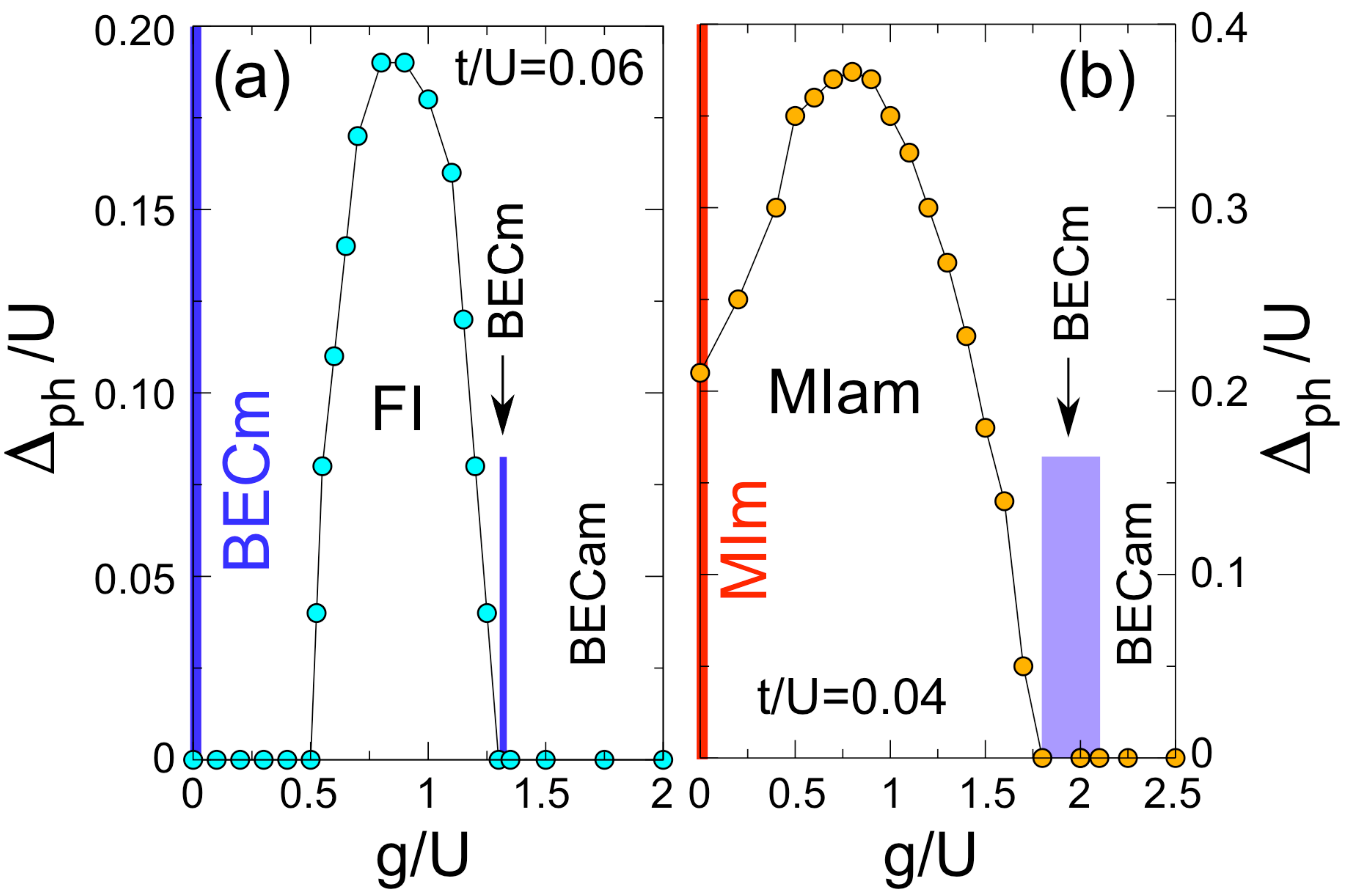}
\caption{Dependence of the particle-hole gap  $\Delta_{ph}$, Eq.~\eqref{phGap},   on the atom-molecule conversion rate $g$ for fixed detuning $\delta/U=-1.2$
and for two hopping values, $t/U=0.06$ (a),  and $t/U=0.04$ (a) (see also Fig.~\ref{fig:1} of the main text).} 
\label{Figure7_SuppMat}
\end{center}
\end{figure}

The p-h gap, $\Delta_{ph}$, is given by the chemical potential width of the insulating phase at $n = n_a + 2n_m = 2$, 
\begin{equation}
\Delta_{ph}(\delta,g) =  \mu_p(\delta,g) - \mu_h(\delta,g)  
\label{phGap}
\end{equation} 
where $\mu_p$ ($\mu_h$) is the critical chemical potential to add a particle (hole) to the incompressible phase.
Fig.~\ref{Figure7_SuppMat} shows the evolution of $\Delta_{ph}$ when $g$ grows. 
In the case $t/U=0.06$ (Fig.~\ref{Figure7_SuppMat}(a)) one observes that a quantum phase transition is driven by the atom-molecule conversion, which opens a $p-h$ gap for a critical $g_c$ value ($g_c/U=0.55$) giving rise to a FI. On the other hand, in the case $t/U=0.04$ (Fig.~\ref{Figure7_SuppMat}(b)), starting in the MIm phase at $g=0$, the already existing gap $\Delta_{ph}$ at $g=0$ is enhanced by the atom-molecule conversion up to $g/U \simeq 1.6$, characterising a Feshbach-stabilized MI of molecules and atoms (MIam).
\begin{figure}[h]
\begin{center}
\includegraphics[width=1 \columnwidth]{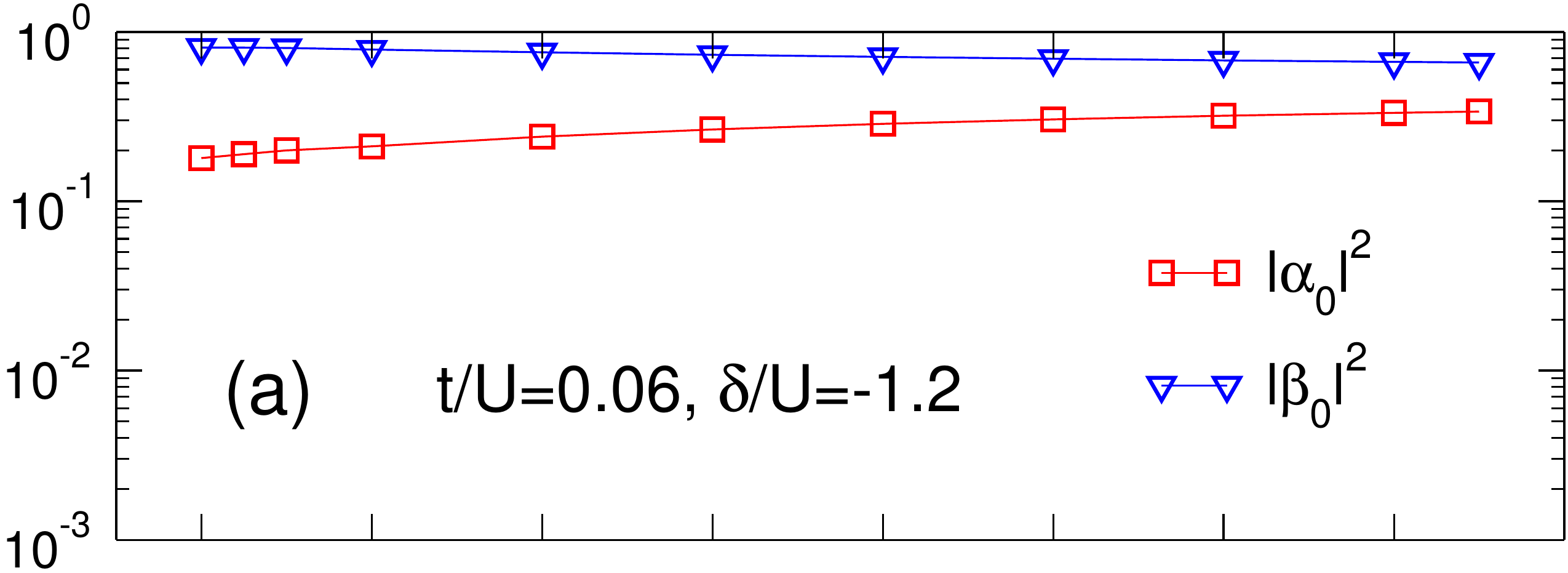} \\
\includegraphics[width=1 \columnwidth]{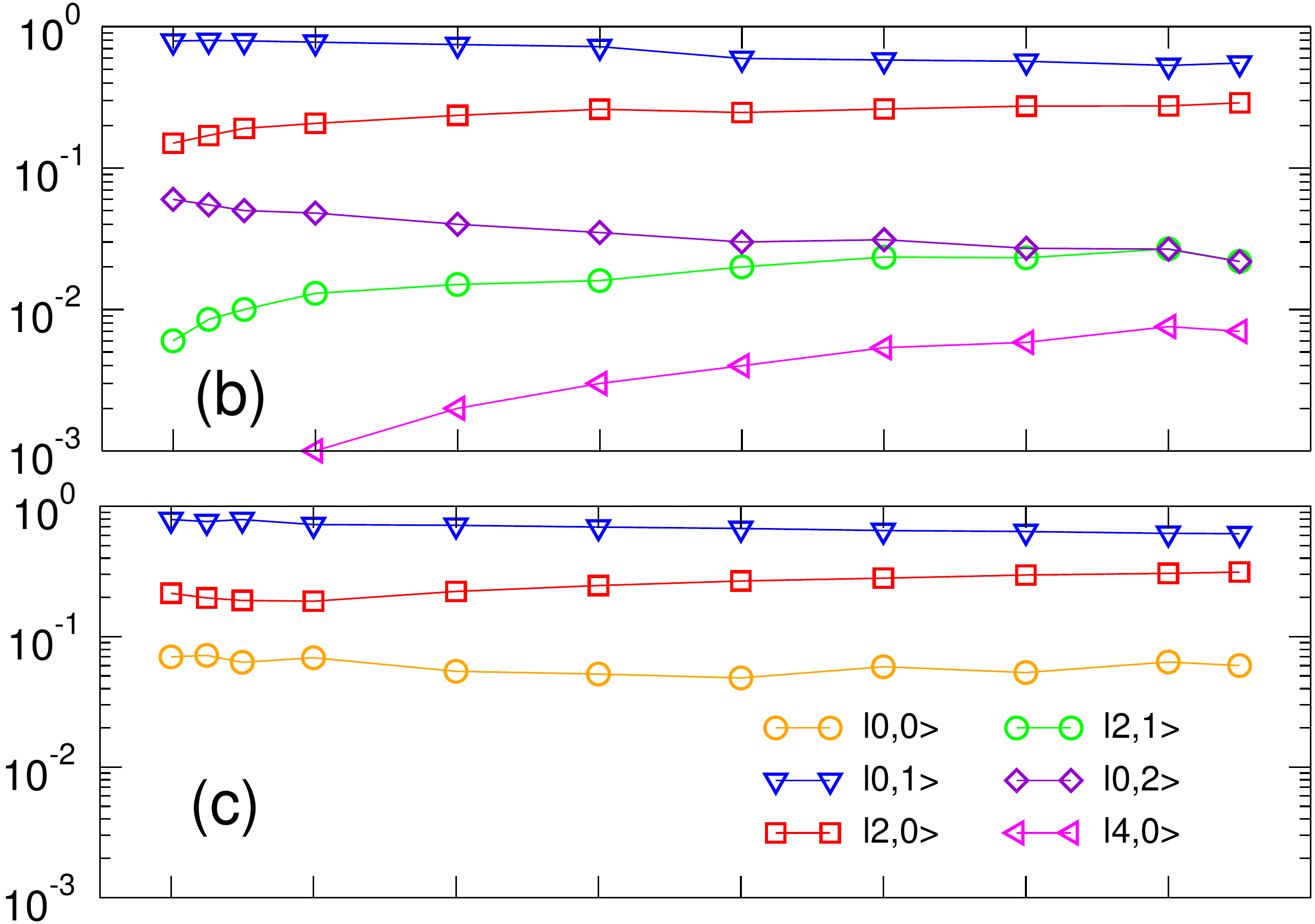} \\
 \hspace{-3. pt}
\includegraphics[width=1 \columnwidth]{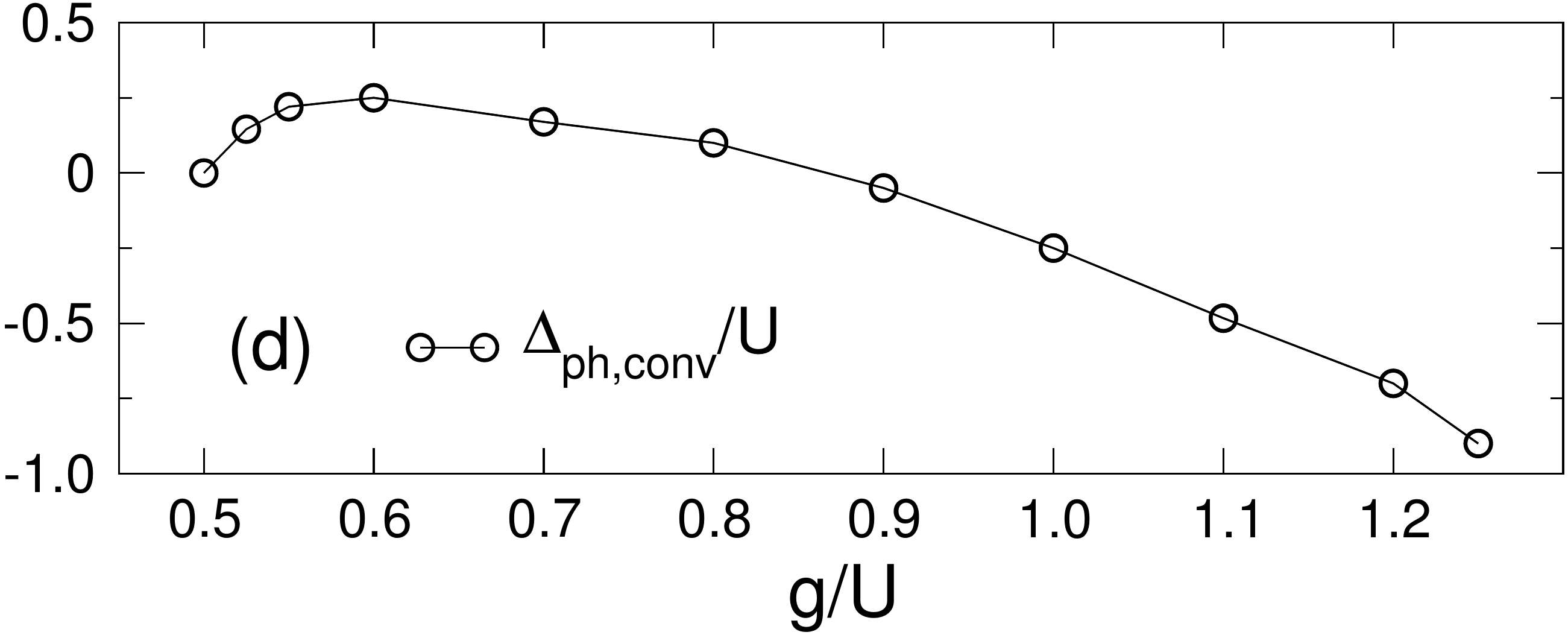}
\caption{$g$-dependence of the wavefunctions of the ground state and excitations, and of their conversion energy, for $t/U=0.06$ and $\delta/U=-1.2$: (a) probabilities $|\alpha_0|^2$ and $|\beta_0|^2$ for the ground-state wave function. Eq.~\eqref{E_ground}; 
(b) probabilities of the ground-state component along a line  $\mu = \mu_p+\epsilon$ with $\epsilon=0.1U_a$; 
(c) probabilities of the ground-state component along a line  $\mu = \mu_h-\epsilon$ with  $\epsilon=0.1U_a$; 
(d) conversion energy of a p-h pair $\Delta_{ph,{\rm conv}}/U_a$, Eq.~\eqref{E_conv}.}
\label{Figure8_SuppMat}
\end{center}
\end{figure}
For both cases, a larger value of $g/U$ has the opposite effect of suppressing the p-h gap, leading to a further quantum phase transition to a molecular BEC (BECm) \textcolor{black}{first, and to an atom-molecule condensate (BECam) for a slightly larger value of $g$. 
 We observe that this non-linear gap evolution is in a very good qualitative agreement with the QMC results Fig.~\ref{fig:1}(c,d) of the main text concerning the full width at half maximum of the momentum distribution. (On the basis of our QMC data we cannot definitely conclude about the presence or absence of a narrow BECm region for the same choice of parameters as in the MFT calculation; nonetheless QMC confirms the existence of a tricritical point at which the three transition lines FI-BECm, FI-BECam and BECm-BECam meet).}

\begin{figure}[h]
\begin{center}
\includegraphics[width=1 \columnwidth]{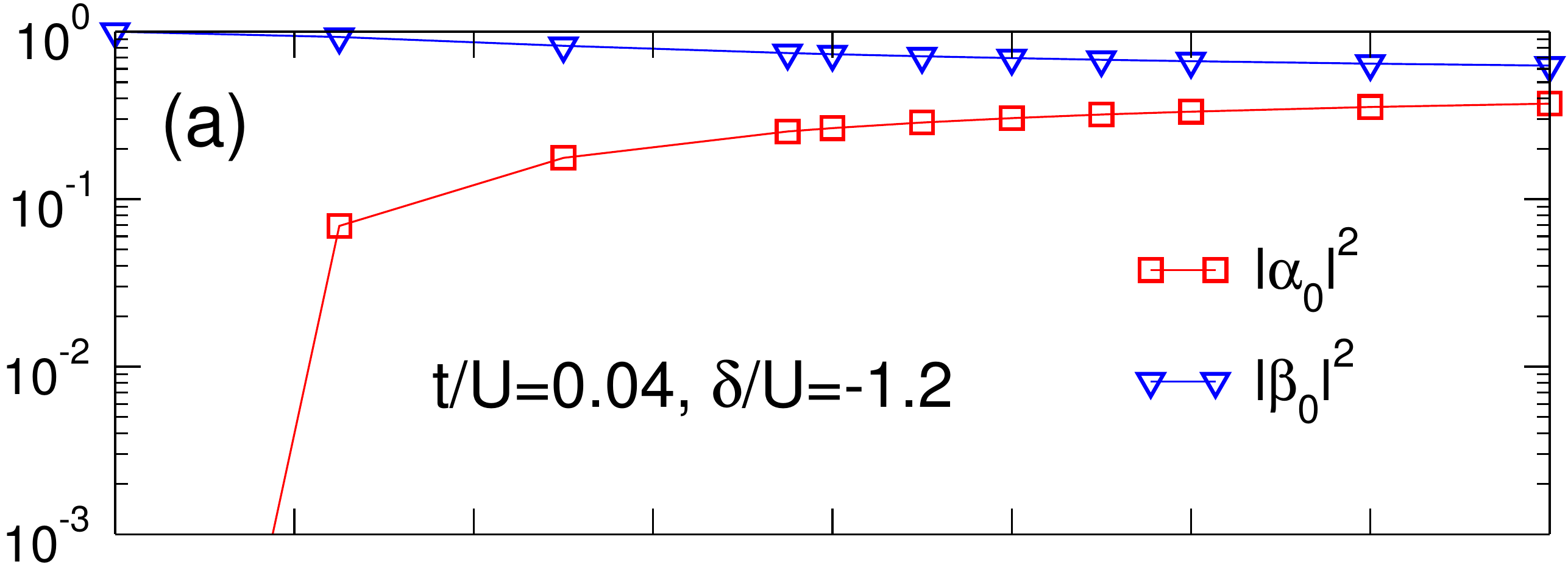} \\
\includegraphics[width=1 \columnwidth]{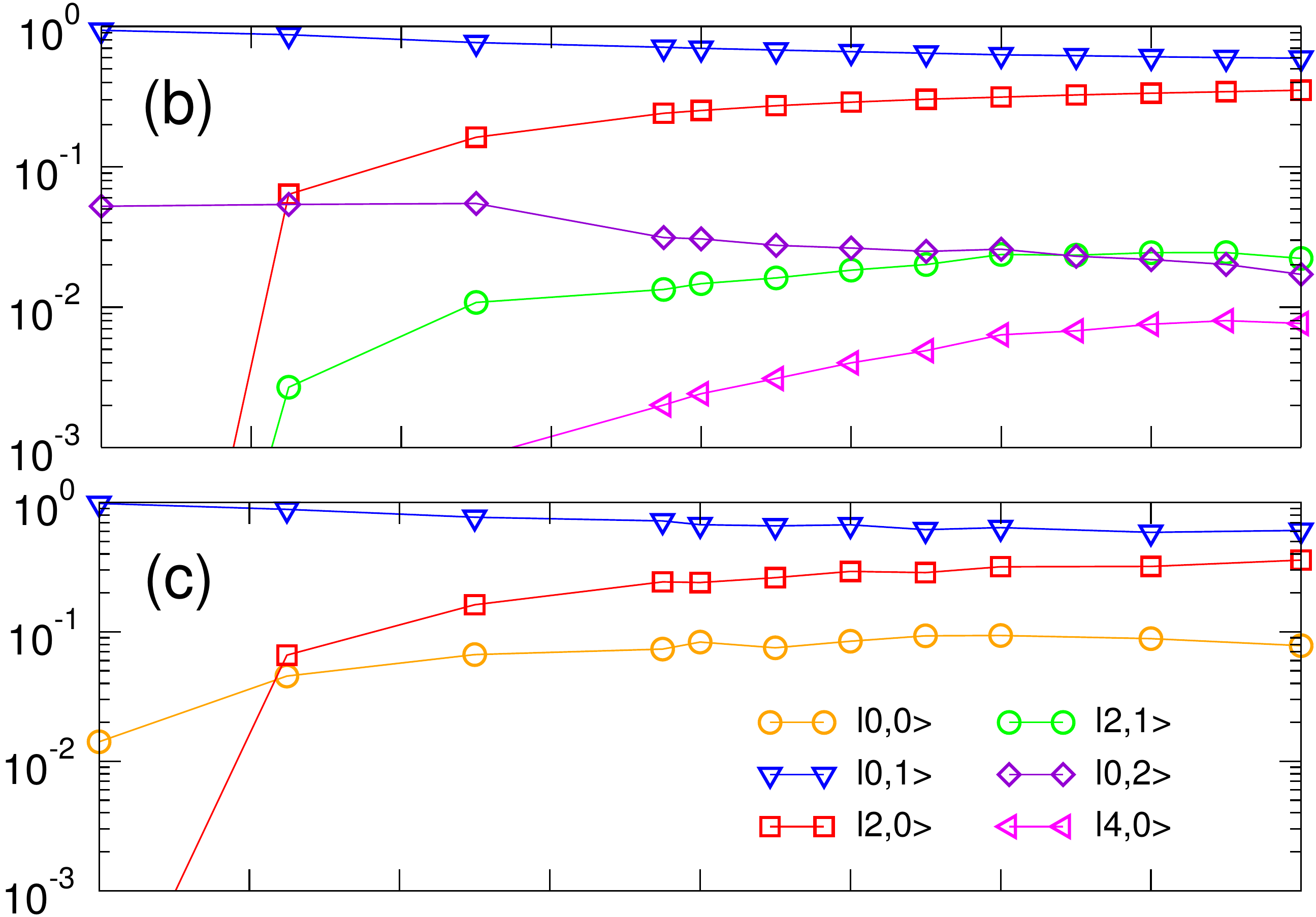} \\
 \hspace{-4 pt}
\includegraphics[width=1 \columnwidth]{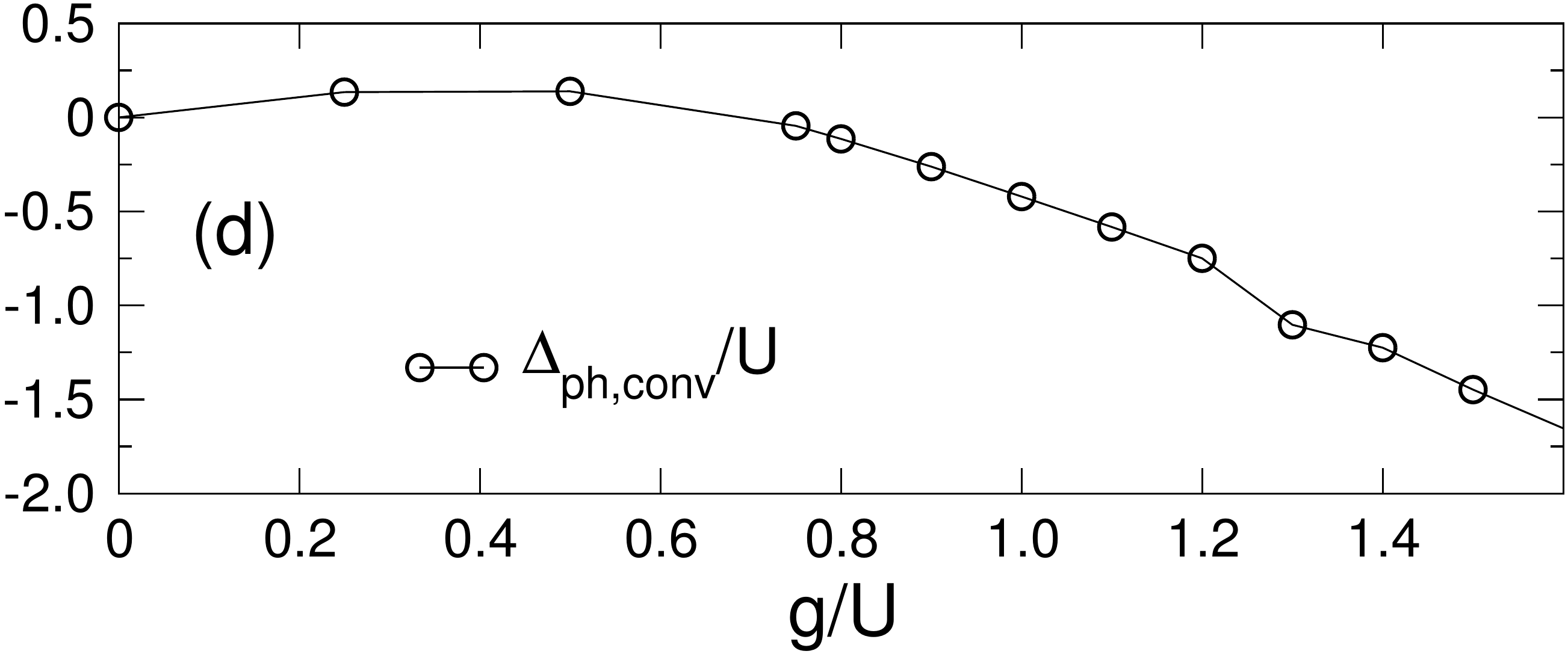}
\caption{$g$-dependence of the wavefunctions of the ground state and excitations, and of their conversion energy, for $t/U=0.04$ and $\delta/U=-1.2$: (a) probabilities $|\alpha_0|^2$ and $|\beta_0|^2$ for the ground-state wave function. Eq.~\eqref{E_ground}; 
(b) probabilities of the ground-state component along a line  $\mu = \mu_p+\epsilon$ with $\epsilon=0.1U_a$; 
(c) probabilities of the ground-state component along a line  $\mu = \mu_h-\epsilon$ with  $\epsilon=0.1U_a$; 
(d) conversion energy of a particle-hole pair $\Delta_{ph,{\rm conv}}/U_a$, Eq.~\eqref{E_conv}.}
\label{Figure9_SuppMat}
\end{center}
\end{figure}

\subsubsection{Wavefunction of the particle and hole excitations}

To gain a detailed microscopic understanding of the role of $g$ in opening the p-h gap of the FI, it is useful to analyze the nature of the particle and hole excitations which are admixed to the ground state at the critical chemical potentials $\mu_p$ and $\mu_h$ respectively. Within mean-field theory, the nature of the excitations can be directly read out of the single-site ground-state wave function
\begin{equation}
|\psi\rangle = \sum_{n_a, n_m} c(n_a,n_m) ~|n_a, n_m\rangle~.
\end{equation}  
Indeed, slightly above the $\mu_p$ chemical potential, namely for $\mu = \mu_p + \epsilon$, one has
\begin{equation}
|\psi\rangle \approx \sqrt{1-\eta_+^2} ~|\psi_0\rangle + \eta_+ |\psi_p\rangle
\end{equation}
where $|\psi_0\rangle$ is the ground state of the insulating phase, $|\psi_p\rangle$ is the wave function of the particle excitation, and $\eta_+ \ll 1$; similarly for $\mu = \mu_h - \epsilon$ one has 
\begin{equation}
|\psi\rangle \approx \sqrt{1-\eta_-^2}~ |\psi_0\rangle + \eta_- |\psi_h\rangle~.
\end{equation}
where again $\eta_- \ll 1$ and $|\psi_h\rangle$ is the wave function for the hole excitation. 
 
 In the presence of an atom-molecule conversion, the insulating ground state with $n = 2$ has the form 
 \begin{equation}
|\psi_0\rangle = \alpha_0 |2,0\rangle + \beta_0 |0,1\rangle 
\label{E_ground}
\end{equation}
where $\alpha_0$ and $\beta_0$ depend on the conversion $g$ as shown in Figs.~\ref{Figure8_SuppMat}(a),~\ref{Figure9_SuppMat}(a). 

When $\mu = \mu_p+\epsilon$, the particle excitation
\begin{equation}
|\psi_p\rangle = \alpha_p |2,1\rangle + \beta_p |0,2\rangle + \gamma_p |4,0 \rangle 
\end{equation}
is admixed with the insulating ground state, while for $\mu = \mu_h-\epsilon$ the hole excitation $|\psi_h\rangle \approx |0,0\rangle$ (or the vacuum) 
 is admixed to the ground state, as shown in Figs.~\ref{Figure8_SuppMat}(b-c),~\ref{Figure9_SuppMat}(b-c). 
This result shows clearly how atom-molecule conversion establishes a fundamental \emph{asymmetry} between the nature of the particle excitation and that of the hole excitation: while the particle excitation is a coherent superposition between an
 ``atom-pair-like" excitation ($|2,1\rangle$ and $|4,0\rangle$) and a ``molecule-like" excitation ($|0,2\rangle$), the hole excitation cannot have such a coherent nature. 
Due to the non-linear dependence of the conversion energy on the particle density, the latter enters in the energy balance of particle hole excitations $E_p + E_h$  (where $E_p$ is the energy cost of the particle excitations and $E_h$ that of the hole excitation). The hole excitation makes the conversion energy vanish, and hence it has a conversion energy cost of $E_{h,{\rm conv}}=2~{\rm Re}(\sqrt{2} \alpha_0^* \beta_0)~g$ starting from the ground state with $n=2$. The particle excitation varies instead the conversion energy by a term $E_{p,{\rm conv}} = -2~{\rm Re}(2 ~\alpha^*_p\beta_p +\sqrt{12}~ \alpha^*_p\gamma_p  -\sqrt{2} ~\alpha^*_0\beta_0)~g$. 
Hence the conversion energy contributes to the p-h excitation energy with a term
\begin{eqnarray}
\Delta_{ph,{\rm conv}} &=& E_{p,{\rm conv}} + E_{h,{\rm conv}} \nonumber \\
&=& 4~{\rm Re}(\sqrt{2}~\alpha^*_0\beta_0 - \alpha^*_p\beta_p -\sqrt{3}  \alpha^*_p\gamma_p)~g~.
\label{E_conv}
\end{eqnarray}

\subsubsection{Conversion energy contribution to the particle-hole gap}

Fig.~\ref{Figure8_SuppMat}(d) shows that, for $t/U=0.06$, $\Delta_{ph,{\rm conv}}$ marks the BECm-FI quantum phase transition, and for sufficiently small $g$ its magnitude is consistent with the $p-h$ gap in Fig.~\ref{Figure7_SuppMat}(a), namely the conversion energy contribution dominates the particle-hole gap. Similarly, for $t/U=0.04$ Fig.~\ref{Figure9_SuppMat}(d) shows that $\Delta_{ph,{\rm conv}}$ is of the order of the $p-h$ gap increment in Fig.~\ref{Figure7_SuppMat}(b) for sufficiently small $g$. For both cases, 
$\Delta_{ph,{\rm conv}}$ turns negative at larger $g$, showing that the atom-molecule conversion plays an opposite, gap-suppressing role, in agreement with the overall $g$-dependence of the gap.

 What is the origin of this complex, dual role of the atom-molecule conversion (gap-opening or gap-stabilizing for small $g$, gap-suppressing for large $g$)? It can be again read out of the structure of the particle and hole excitations. For small $g$, $\gamma_p$ is negligible, and ${\rm Re}(\alpha_0 \beta_0) > {\rm Re}(\alpha_p^* \beta_p)$: the atom-molecule admixture in the ground state is more significant than in the particle excitation state. This latter aspect is \emph{generic}, because the overall atom-molecule detuning (combining the explicit $\delta$ detuning as well as the interaction and kinetic energy difference) is more significant at higher fillings, namely it is stronger in the particle-excitation state than in the ground state. Hence for small $g$ the atom-molecule conversion is more efficient in the $n=2$ ground state than in the particle-excitation state, and therefore it drives the mechanism behind the gap opening/gap enhancement. 
 
 Nonetheless an increase in $g$ bridges this difference between insulating ground state and particle excitation. Indeed the atom-molecule conversion $g$ is all the more effective the larger the filling (due to the boson-enhancement factor), and when $g\sim U, \delta$ one reaches the condition $\alpha_p = \beta_p$ at a lower $g$ then that required for the condition $\alpha_0 = \beta_0$ (not shown in the figures).  As a consequence, ${\rm Re}(\alpha_p^* \beta_p)$ grows faster than ${\rm Re}(\alpha_0 \beta_0)$; moreover the appearance of a non-negligible $\gamma_p$ introduces a further reduction term in $\Delta_{ph,{\rm conv}}$, and these two aspects together drive the p-h towards zero. The lowering of the energy cost of the particle excitation comes from the fact that the atom-molecule conversion favors clustering of atoms and molecules: this is again due to the boson enhancement factor, which introduces an explicit density dependence of the conversion energy. As a result, within MFT in the grand-canonical ensemble a large $g$ is found to drive a quantum phase transition from an insulating phase with $n=2$ to a BECm with $n>2$.
 
 \subsection{$^{87}$Rb close to its 414-G Feshbach resonance}

 \subsubsection{Dependence of the particle-hole gap on conversion rate and detuning}
 \label{s.dDeltadg}
\begin{figure}[ht!]
\begin{center}
\includegraphics[width=1 \columnwidth]{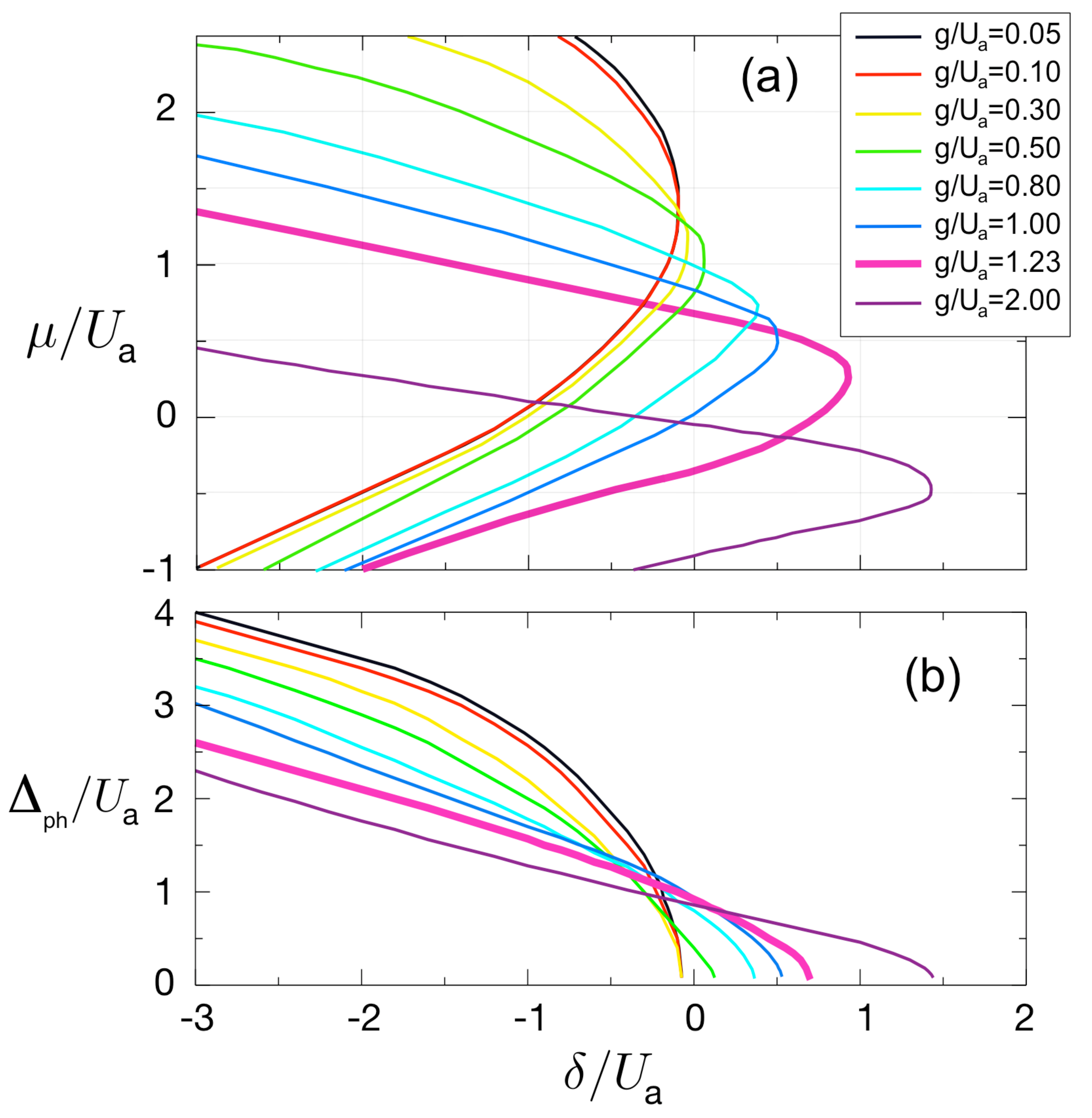}
\caption{ Dependence of the insulating region (a) and of the particle-hole gap  
$\Delta_{ph}/U_a$ (b) on the atom-molecule conversion $g/U_a$ for 
 $^{87}$Rb; all other parameters as in Fig.~\ref{fig:3} of the main text.} 
\label{Figure10_SuppMat}
\end{center}
\end{figure}

\begin{figure}[ht!]
\begin{center}
\includegraphics[width=\columnwidth]{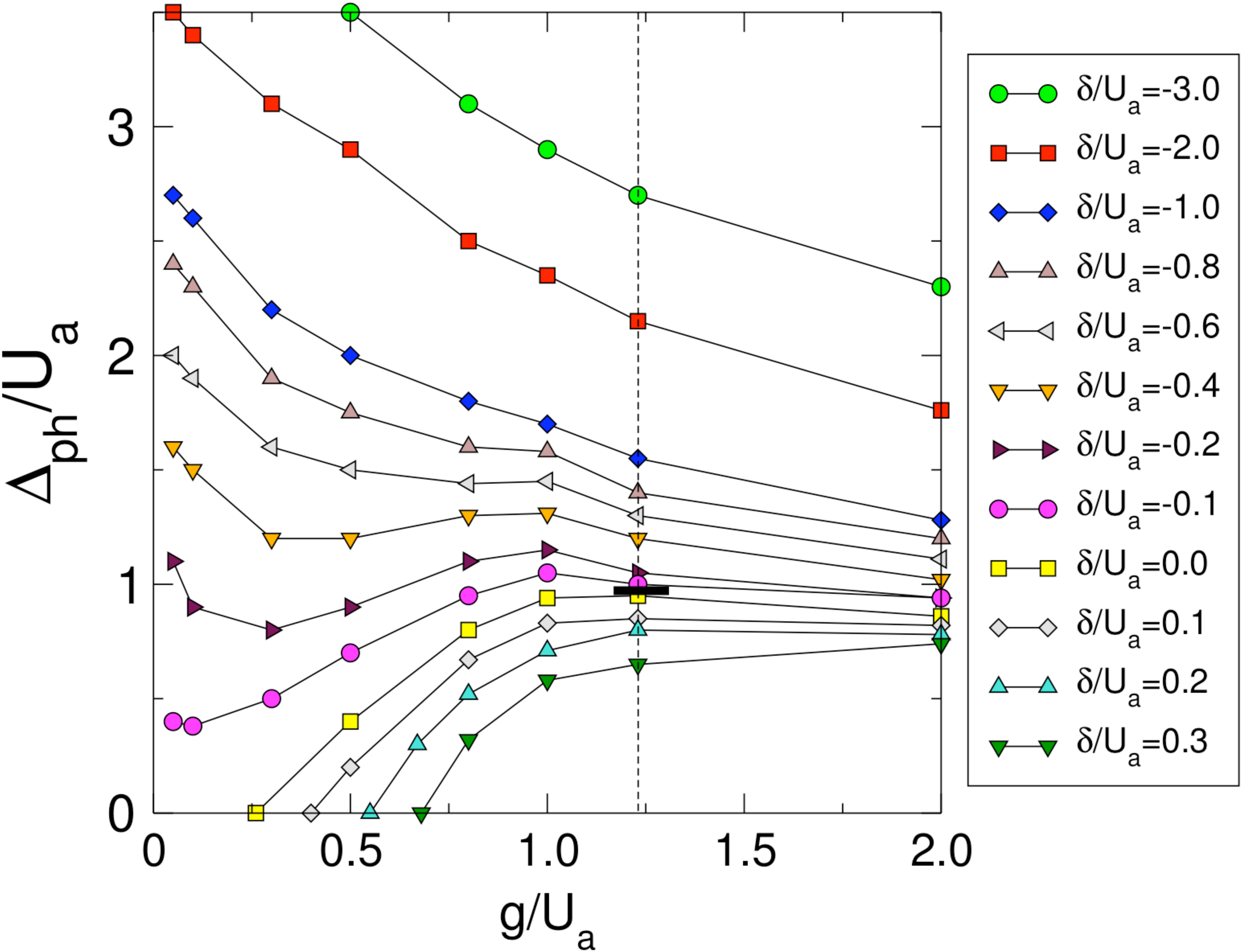}
\caption{ Evolution of the particle-hole gap as a function of the atom-molecule conversion $g/U_a$ for several values of the detuning $\delta/U_a$; other parameters as in 
Fig.~\ref{fig:3} of the main text. The vertical dashed line marks the value $g/U_a = 1.23$ of $^{87}$Rb at its 414 G resonance. The FI region for this resonance appears for $\delta/U_a>\delta_c/U_a = -0.05$ (as marked by the think horizontal line), at which the derivative $\partial\Delta_{ph}/{\partial g}$ at $g/U_a = 1.23$ goes from negative to positive.} 
\label{Figure11_SuppMat}
\end{center}
\end{figure}

 We now focus on the phase diagram of $^{87}$Rb, Fig.~\ref{fig:3} of the main text.
 To understand how the atom-molecule conversion acts in opening a p-h gap on the atomic side of the resonance, 
 it is instructive to monitor the evolution of the insulating phase in the phase diagram of 
 Fig.~\ref{fig:3}(a) of the main text upon changing $g$. As discussed in the main text and later in Sec.~\ref{s.modulation}, a continuous tuning of $g$ between zero and its intrinsic value for the chosen Feshbach resonance is achievable 
 in the experiment via periodic modulation of the atom-molecule detuning, controlled by the magnetic field.  
 Fig.~\ref{Figure10_SuppMat}(a) shows the evolution of the insulating lobe at $n = n_a + 2n_m = 2$ when $g$ grows. 
 The evolution with $\delta$ of the gap $\Delta_{ph}$, Eq.~\eqref{phGap}, for different values of $g$ is further plotted in Fig.~\ref{Figure10_SuppMat}(b), while the $g$-dependence for different values of $\delta$ is shown in Fig.~\ref{Figure11_SuppMat}. One can clearly distinguish two separate regimes in the $g$-dependence of the gap: I) for sufficiently negative $\delta$ the gap decreases monotonically with $g$, a behavior well understood in terms of the renormalisation effect that (virtual) atoms have on molecule-molecule interactions. At the lowest order in perturbation theory, the molecule-molecule interaction is decreased from $U_m$ by a term of order $g^4/\delta^3$; II) on the other hand, for $\delta\gtrsim 0$ the gap is monotonically \emph{increasing} with $g$, starting from zero at $g=0$. This latter behavior stems from the non-perturbative role of the coherent atom-molecule conversion, and we identify it as the distinctive feature of a FI. In particular, for the case of $^{87}$Rb ($g/U_a$ = 1.23) one observes that the gap changes from being a decreasing function of $g$ to being an increasing 
 function of $g$ for $\delta \approx \delta_c = -0.05U_a$, and that this value of $\delta$ marks the separatrix between curves $\Delta_{ph}(g;\delta)$ which vanish when $g\to 0$ and those who do not. Hence this value marks the crossover from Mott insulating to FI behavior, as indicated in the phase diagram of Fig.~\ref{fig:3}(a) of the main text. 

 \subsubsection{Wavefunction of the particle and hole excitations}

\begin{figure}[h]
\begin{center}
\includegraphics[width=1 \columnwidth]{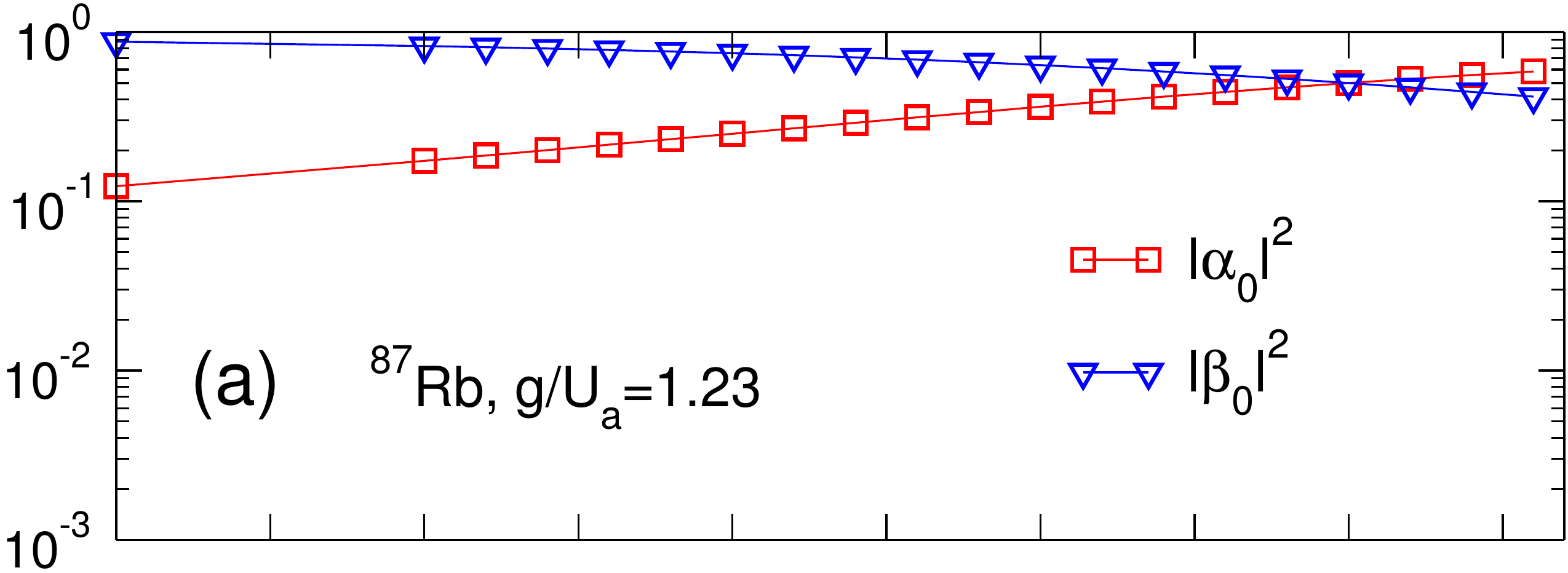} \\
\includegraphics[width=1 \columnwidth]{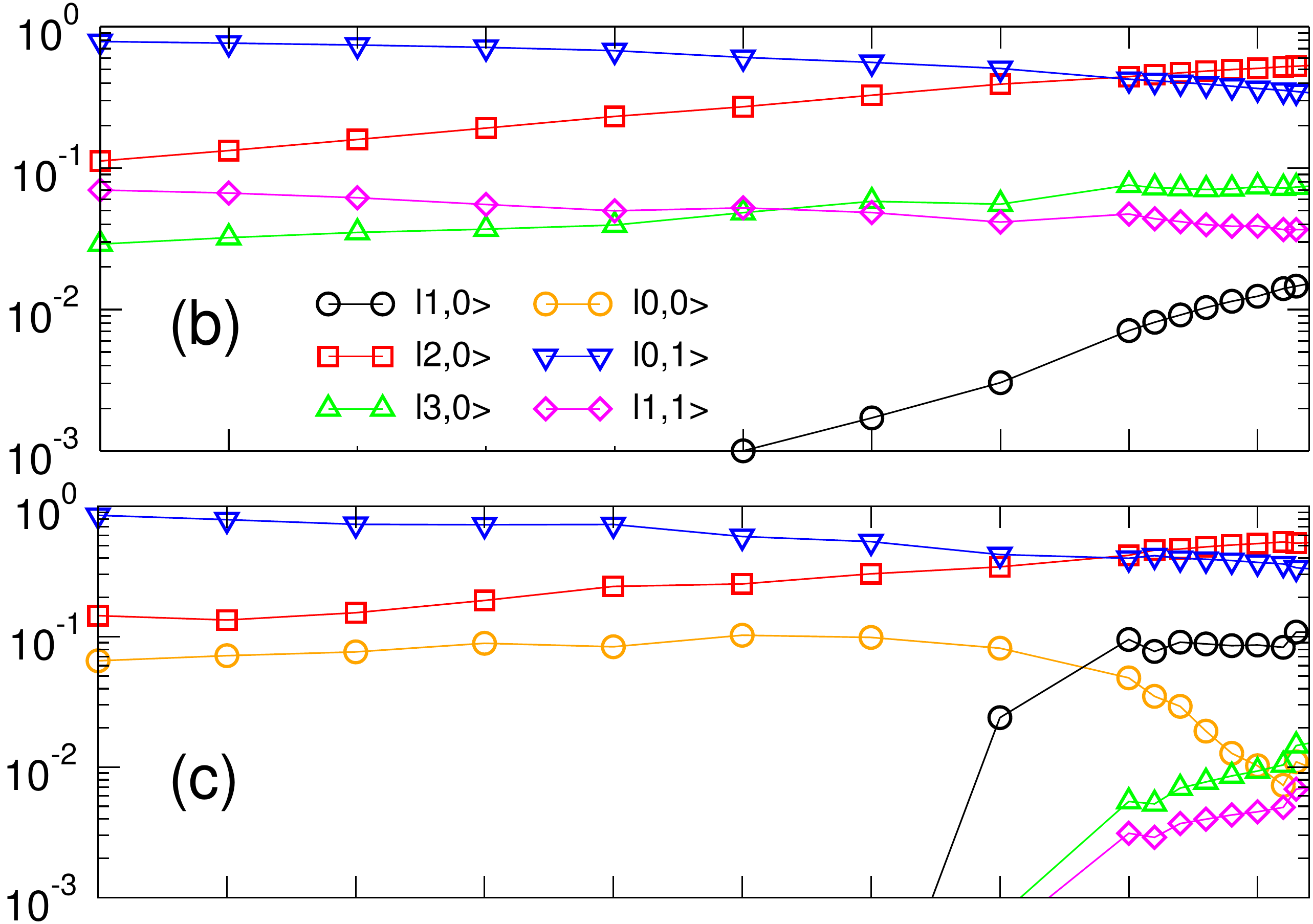} \\
\includegraphics[width=1 \columnwidth]{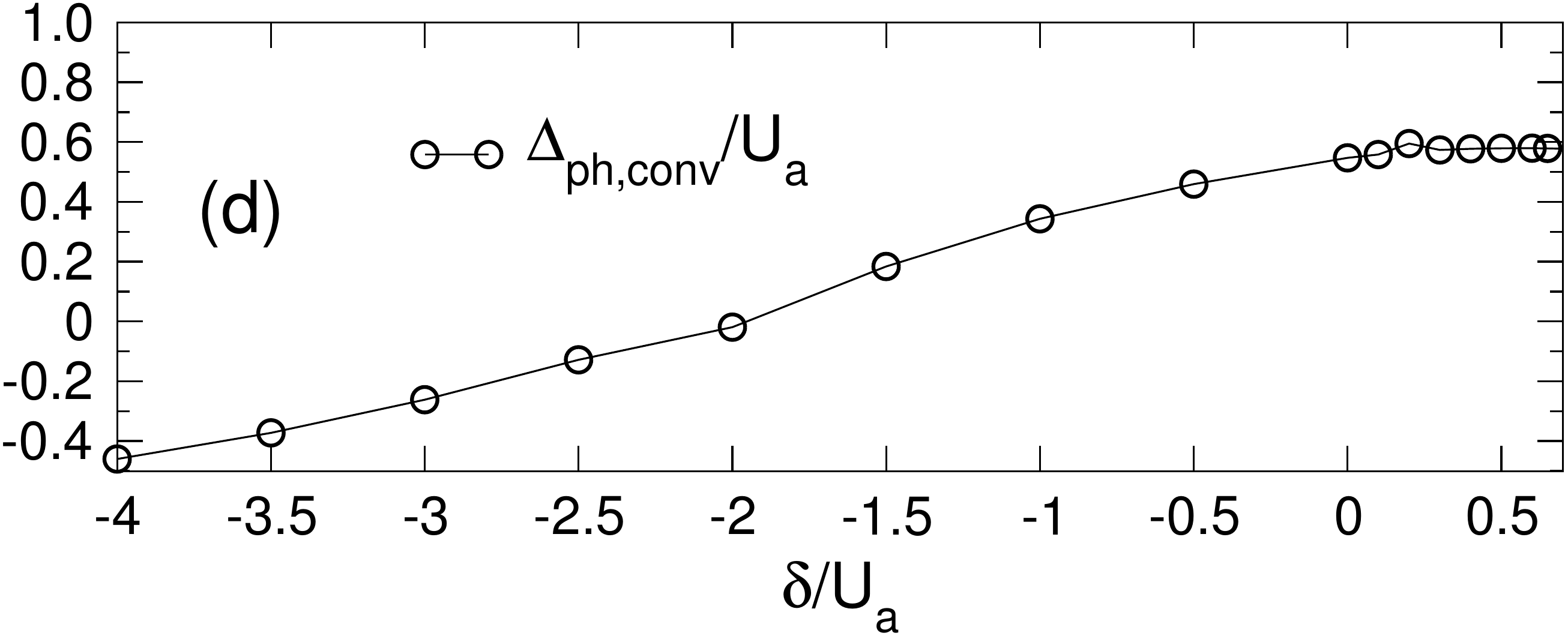}
\caption{ (a) Probabilities $|\alpha_0|^2$ and $|\beta_0|^2$ for the ground-state wave function, Eq.~\eqref{E_ground}; 
(b) probabilities of the ground-state component along a line  $\mu = \mu_p+\epsilon$ with $\epsilon=0.1U_a$; 
(c) probabilities of the ground-state component along a line  $\mu = \mu_h-\epsilon$ with  $\epsilon=0.1U_a$; 
(d) conversion energy of a particle-hole pair $\Delta_{ph,{\rm conv}}/U_a$, Eq.~\eqref{e.conv}.}
\label{Figure12_SuppMat}
\end{center}
\end{figure}

Similarly to the symmetric atom-molecule mixture, the nature of the particle and hole excitations give a microscopic understanding of the role of $g$ in opening the p-h gap of the FI. In the insulating phases (MIam and FI),  the insulating ground state has the form
$|\psi_0\rangle = \alpha_0 |2,0\rangle + \beta_0 |0,1\rangle$, 
where $\alpha_0$ and $\beta_0$ depend on the detuning $\delta$, and in particular $\alpha_0 = \beta_0 = 1/\sqrt{2}$ for $\delta = 0$, as shown in Fig.~\ref{Figure12_SuppMat}(a). 
As shown in Fig.~\ref{Figure12_SuppMat}(b-c), when $\mu = \mu_p+\epsilon$ the particle excitation
\begin{equation}
|\psi_p\rangle = \alpha'_p |3,0\rangle + \beta'_p |1,1\rangle 
\end{equation}
is admixed with the insulating ground state, while for $\mu = \mu_h-\epsilon$ the hole excitation is admixed to the ground state, and in particular it is predominantly a single hole, $|\psi_h\rangle \approx |1,0\rangle$, for $\delta > 0$, while it is predominantly a double hole (or the vacuum) $|\psi_h\rangle \approx |0,0\rangle$ for $\delta < 0$.  
Similarly to the symmetric case,  the  atom-molecule conversion establishes a fundamental \emph{asymmetry} between the nature of the particle excitation and that of the hole excitation: while the particle excitation is a coherent superposition between an "atom-like" excitation ($|3,0\rangle$) and a "molecule-like" excitation ($|1,1\rangle$), the hole excitation does not possess such a coherent nature and, for instance, it is predominantly atom-like ($|1,0\rangle$) for $\delta>0$. The hole excitation makes the conversion energy has again a conversion energy cost of $E_{h,{\rm conv}}=2~{\rm Re}(\sqrt{2} \alpha_0^* \beta_0)~g$ starting from the ground state with $n=2$. The conversion energy variation due to the particle excitation is instead $E_{p,{\rm conv}} = -2~{\rm Re}(\sqrt{6} ~\alpha'^*_p\beta'_p-\sqrt{2} ~\alpha^*_0\beta_0)~g$. As a result the conversion energy contribution to the p-h excitation energy reads
\begin{eqnarray}
\Delta_{ph,{\rm conv}} &=& E_{p,{\rm conv}} + E_{h,{\rm conv}} \nonumber \\
&=& 2\sqrt{2}~{\rm Re}(2~\alpha^*_0\beta_0-\sqrt{3}~\alpha'^*_p\beta'_p)~g~.
\label{e.conv}
\end{eqnarray}
 Fig~\ref{Figure12_SuppMat}(d) shows that this quantity is positive and almost constant all over the FI regime $-0.05\lesssim \delta/U_a \lesssim 0.7$. Therefore it remains positive even on the BECam side -- at least close to the crossover to FI -- marking a regime of Feshbach-enhanced MI. On the other hand,  $\Delta_{ph,{\rm conv}}$ becomes negative sufficiently far from the resonance on the molecular side, marking instead the conventional regime of Feshbach-suppressed MI. 

 \subsubsection{Opening of the particle-hole gap in the Feshbach insulator}
 
 The above data clearly explain the mechanism for the opening of a p-h gap in the FI regime. When coming from the BECam side at $\delta > 0.7 ~U_a$, the conversion term introduces a positive energy cost in the p-h excitation, Eq.~\eqref{e.conv}, which depends on the atom-molecule coherences $\alpha^*_0\beta_0$ and $\alpha'^*_p\beta'_p$. The latter are controlled by the detuning $\delta$, and become maximal around resonance.  As the prefactor of $g$ in Eq.~\eqref{e.conv} is  upper-bounded, it is necessary that $g$ be sufficiently large (namely, the Feshbach resonance be sufficiently broad) for the conversion energy cost to outgrow the kinetic-energy gain of a p-h excitation and to open a gap. Figs.~\ref{Figure10_SuppMat} and  \ref{Figure11_SuppMat} show indeed that for $\delta\geq\delta_c=-0.05U_a$, namely on the atomic side of the resonance, $g$ needs to exceed a $\delta-$dependent critical value $g_c(\delta)$ for a p-h gap to open. $g_c$ is actually nonzero for \emph{any} $\delta\geq \delta_c$. This means that a FI phase can be realized by $^{87}$Rb at the 414 G resonance because the width of the resonance not only is sufficiently narrow to avoid the collapse of the system, but it is also sufficiently large, namely $g>g_c(\delta_c)$. The interparticle repulsion, which drives the opening of a p-h gap for $\delta \leq \delta_c$, has mostly a ``spectator" role for the FI, stabilizing the density of the system around the value $n=2$, and hence preventing the system from collapsing. 
 

\section{7. Atom-molecule coherence from momentum-noise correlations} 

The FI is characterized by a strong atom-molecule coherence 
${\cal C} = \langle a^2 m^{\dagger} \rangle$. Can one measure the coherence ${\cal C}$ in an experiment?

A possibility comes from the analysis of the correlations between the noise of atomic and molecular momentum distributions. The momentum distribution of atoms and molecules can be measured simultaneously using Stern-Gerlach separation during time of flight \cite{SM_Herbigetal2003}. This leads us to consider the correlation function
\begin{equation}
G^{(2a,m)}({\bm k}, {\bm k'}) = \langle n^2_a({\bm k}) ~n_m({\bm k}') \rangle 
\end{equation}
where $n_a({\bm k})$ and $n_b({\bm k})$ are the momentum distribution of atoms and molecules respectively.   

We assume for simplicity that the state of the system factorises between sites, namely 
\begin{equation}
|\Psi\rangle = \bigotimes_{i=1}^N \left( \cos(\theta/2) |2,0\rangle + \sin(\theta/2) |0,1\rangle \right)
\label{e.state}
\end{equation}
where the state is expressed in the $|n_a , n_m\rangle$ basis. 

Then we readily obtain that 
\begin{align}
G^{(2a,m)}({\bm k}, {\bm k'}) = ~~~~~~~~~~~~~~~~~~~~~~~~~~~~~~~~~~~~~~~~~~~~~~~~~ \nonumber \\
2 \langle n^2_a \rangle \langle n_m \rangle + \frac{n_m n_a}{N} + \frac{|{\cal C}|^2}{N} \delta_{2{\bm k}-{\bm k'}, {\bm K}} + {\cal O}\left(\frac{1}{N^2}\right) 
\end{align}
where $n_a$, $n_m$ are the average atomic and molecular densities, ${\bm K}$ is a reciprocal lattice vector, and ${\cal C}=(\sin\theta)/2$. 

Introducing then the function
\begin{equation}
g^{(2)}_{2a,m}({\bm q}) = \frac{ \sum_{\bm k}  G^{(2a,m)}({\bm k}, 2{\bm k}+{\bm q}) }{\sum_{\bm k} \langle n_a^2({\bm k}) \rangle \langle n_m(2{\bm k}+{\bm q})\rangle}
\end{equation} 
we find that, for the factorized state of Eq.~\eqref{e.state}
\begin{equation}
g^{(2)}_{2a,m}({\bm q}) = 1 + \frac{ |{\cal C}|^2}{2 n_a^2 n_m} \frac{1}{N} \delta_{{\bm q},{\bm K}} + {\cal O}\left(\frac{1}{N^2}\right)
\end{equation}
namely this function exhibits a series of \emph{bunching peaks} describing the reciprocal lattice, and whose height is proportional to the square of the atom-molecule coherence. Hence the noise correlations in the fluctuations of the atomic and molecular momentum distributions reveal the presence of atom-molecule coherence. 
Yet an important limitation of this approach is that the bunching peaks are suppressed like $1/N$. This means that the bunching signal can only be seen on relatively \emph{small samples} -- as it is a factor of $N$ weaker than the typical bunching signal of a Mott insulator in a very deep optical lattice, as detected \emph{e.g.} in \cite{SM_Follingetal2005}. 


\section{8. Control of the atom-molecule conversion rate via modulation of the magnetic field} 
\label{s.modulation}

 In this section we show how to continuously reduce the value of the atom-molecule conversion $g$ from its intrinsic value associated with the chosen Feshbach resonance down to zero, making use of the periodic modulation of the magnetic field, which controls the atom-molecule detuning $\delta$. 

 We start our discussion by considering a simple single-site problem, in which the atom-molecule detuning is driven periodically at a frequency $\omega$:
 \begin{equation}
{\cal H} = -g \left[  (a^{\dagger})^2 m + {\rm h.c.} \right] + [\delta_0 + \delta_1 \cos(\omega t)] ~m^{\dagger} m + ....
 \end{equation}
 The $...$ stand for terms diagonal in the atomic and molecular densities, which will not be affected by the subsequent manipulations.  

Consider then the solution $|\psi(t)\rangle$ to the time-dependent Schr\"odinger equation governed by the Hamiltonian ${\cal H}$. 
Introducing the unitary operator
\begin{equation}
U(t) = \exp\left [i \frac{\delta_1}{\hbar \omega} \sin(\omega t)  m^{\dagger} m \right]~,
\end{equation}
the state $|\tilde\psi(t)\rangle = U(t) |\psi\rangle$ evolves according to the effective Hamiltonian
 $\tilde{\cal H} = U {\cal H} U^{\dagger} + i\hbar \left( \frac{d}{dt}U \right) U^{\dagger}$ which takes the form 
 \begin{equation}
\tilde {\cal H} = -g(t) \left[  (a^{\dagger})^2 m + {\rm h.c.} \right] + \delta_0  ~m^{\dagger} m +....
 \end{equation}
where 
$$ g(t) = g \exp\left [i \frac{\delta_1}{\hbar \omega} \sin(\omega t)\right] ~.$$
Assuming that $\omega$ is not resonant with any transition in the system, and in particular $\omega \ll \delta_0/\hbar$, the oscillating part of the $g(t)$ coupling constant has little effect on the dynamics, so that it is justified to approximate $g(t)$ with its time average:
$$ g(t) \approx g_{\rm eff} = \frac{1}{T} \int_0^T dt ~ g(t)  = g J_0\left(\delta_1/\hbar \omega \right)$$
where $J_0$ is the zero-th Bessel function of the first kind. This function goes from 1 to 0 when its argument goes from 0 to $\approx 2.4$. Therefore a weak periodic modulation with amplitude $\delta_1$ going from to zero to a value $\approx 2.4 ~\hbar \omega \ll \delta_0$ allows to continuously \emph{reduce} the strength of the atom-molecule coupling from its bare value down to zero. Notice that past experiments investigating the periodic modulation of the magnetic field \cite{SM_Thompsonetal2005} have rather explored the resonant regime $\hbar \omega \approx \delta_0$ to drive the association of molecules away from the resonance, but this implies that the regime of weak driving at low frequency is equally accessible. 

The previous discussion generalizes immediately to the full Hamiltonian of the problem, Eq.(1-3) of the main text. The unitary transformation generalizes to 
\begin{equation}
U(t) = \exp\left [i \frac{\delta_1}{\hbar \omega} \sin(\omega t)  \sum_i m_i^{\dagger} m_i \right]~,
\end{equation}
and it commutes with all the terms of Hamiltonian (conserving the number of molecules) except the atom-molecule conversion one. Hence the continuous control on the $g$ coupling is generally possible in the full many-body setting.



\begin{thebibliography}{}
\bibitem{footnote:H} The restriction to a single band is justified as long as the atom-molecule conversion term $g$ is significantly smaller than the band gap. This is verified in all our calculations.
\bibitem{footnote:oneD} The FI phase exists in $D=1$ as well, but it has not been discussed in the previous literature.
\bibitem{footnote:resonance} The FI appears for $\delta/U \lesssim -1$, namely on the molecular side of the (on-site) resonance. This is easily understood by looking at the case $g=0$: in the system at hand the relevant states coupled by the $g$ terms are a BECa and a BECm, whose degeneracy is not reached for $\delta=0$ but for $\delta<0$ ($\delta=0$ would be the resonance condition between atomic and molecular MI). The shift of the resonance to negative $\delta$ stems from the mismatch in kinetic energy between atoms and molecules (in the case at hand, even if $t_a = t_m$ the molecules are - at most - half the number the atoms); to stabilize a molecular state, the increase in kinetic energy has then to be compensated by a negative detuning.
\bibitem{footnote:recoil} \textcolor{black}{ The recoil energy is defined as $E_r = h^2/2m\lambda^2$ for particles of mass $m$ in a laser wave of wavelength $\lambda$; we chose $\lambda = 830$ nm.}
\bibitem{footnote:nonmonotonic} In Fig.~\ref{fig:4} it is interesting to observe the non-monotonic behavior of the atomic coherence as $\delta$ is reduced towards the resonance: the coherence peak is enhanced first -- as expected from the reduction of the effective scattering length -- but then it is suppressed, owing to the localizing effect of the resonant atom-molecule conversion term. Moreover, due to the first-order nature of the transition, an adiabatic sweep in $\delta$ should lead to hysteresis, namely to a history-dependent form of the momentum distribution depending on whether the detuning is swept across the transition coming from the molecular or the atomic side.
\bibitem{footnote:normalization} The normalization of the momentum distribution is such that $n(k)d^D k$ gives the number of particles in the infinitesimal volume $d^D k$ of momentum space.
\end{thebibliography}
\end{document}